%% file: 129.tex
\documentclass[runningheads]{llncs}

\usepackage{tabularx}
\usepackage{booktabs} 
\usepackage{graphicx}
\usepackage{multirow}
\usepackage{balance}  

\usepackage{algorithm}
\usepackage{algpseudocode}
\usepackage{amsmath}
\usepackage{mathtools}
\usepackage{amsmath}
\usepackage{amssymb}
\usepackage{mathrsfs}
\usepackage{subcaption}
\usepackage{hyperref}
\usepackage[all]{hypcap}

\floatname{algorithm}{\textbf{Algorithm}}

\begin{document}

\mainmatter  

\title{Matching Consecutive Subpatterns Over Streaming Time Series}

\titlerunning{Matching Consecutive Subpatterns Over Streaming Time Series}

\author{Rong Kang\inst{1,2}\and
	Chen Wang\inst{1,2} \and
	Peng Wang\inst{3} \and
	Yuting Ding\inst{1,2} \and
	Jianmin Wang\inst{1,2}}
\authorrunning{R. Kang et al.}
%
\institute{School of Software, Tsinghua University, Beijing, China\\ 
		\and
		National Engineering Laboratory for Big Data Software, China\\
			\email{\{kr11, dingyt16\}@mails.tsinghua.edu.cn}\\ 
		\email{\{wang\_chen, jimwang\}@tsinghua.edu.cn}
	\and
	School of Computer Science, Fudan University, Shanghai, China \\
	\email{pengwang5@fudan.edu.cn}}

\maketitle

\begin{abstract}
Pattern matching of streaming time series with lower latency under limited computing resource comes to a critical problem, especially as the growth of Industry 4.0 and Industry Internet of Things. However, against traditional single pattern matching model, a pattern may contain multiple subpatterns representing different physical meanings in the real world.
Hence, we formulate a new problem, called ``consecutive subpatterns matching'', which allows users to specify a pattern containing several consecutive subpatterns with various specified thresholds. 
We propose a novel representation Equal-Length Block (ELB) together with two efficient implementations, which work very well under all $ L_p $-Norms without false dismissals. 
Extensive experiments are performed on synthetic and real-world datasets to illustrate that our approach outperforms the brute-force method and MSM, a multi-step filter mechanism over the multi-scaled representation  by orders of magnitude. 

\end{abstract}

\input{sec1-introduction}
\input{sec2-related-work}
\input{sec3-problem-definition}

\input{sec4-elb}
\input{sec5-experiment}

\input{sec6-conclusion}

\bibliographystyle{abbrv} 
\bibliography{subpattern}

\end{document}

%% file: sec1-introduction.tex
\section{Introduction}
\label{sec:introduction}

Time series are widely available in diverse application areas, such as Healthcare \cite{wei_atomic_2005}, financial data analysis \cite{wu_online_2004} and sensor network monitoring \cite{zhu_efficient_2003}, and they turn the interests on spanning from developing time series database \cite{jensen_time_2017}.
In recent years, the rampant growth of Industry 4.0 and Industry Internet of Things, especially the development of intelligent control and fault prevention to complex equipment on the edge, urges more challenging demands to process and analyze streaming time series from industrial sensors with low latency under limited computing resource \cite{zhao_adaptive_2014}.

As a typical workload, similarity matching over streaming time series has been widely studied for fault detection, pattern identification and trend prediction, where accuracy and efficiency are the two most important measurements to matching algorithms \cite{lian_similarity_2007}. Given a single or a set of patterns and a pre-defined threshold, traditional similarity matching algorithms aim to find matched subsequences over incoming streaming time series, between which the distance is less than the threshold. 
However, in certain scenarios, the single threshold pattern model is not expressive enough to satisfy the similarity measurement requirements.
Let us consider the following example.

\begin{figure}[!htb]
	\centering
	\begin{minipage}[t]{0.48\textwidth}
	\includegraphics[width=\textwidth]{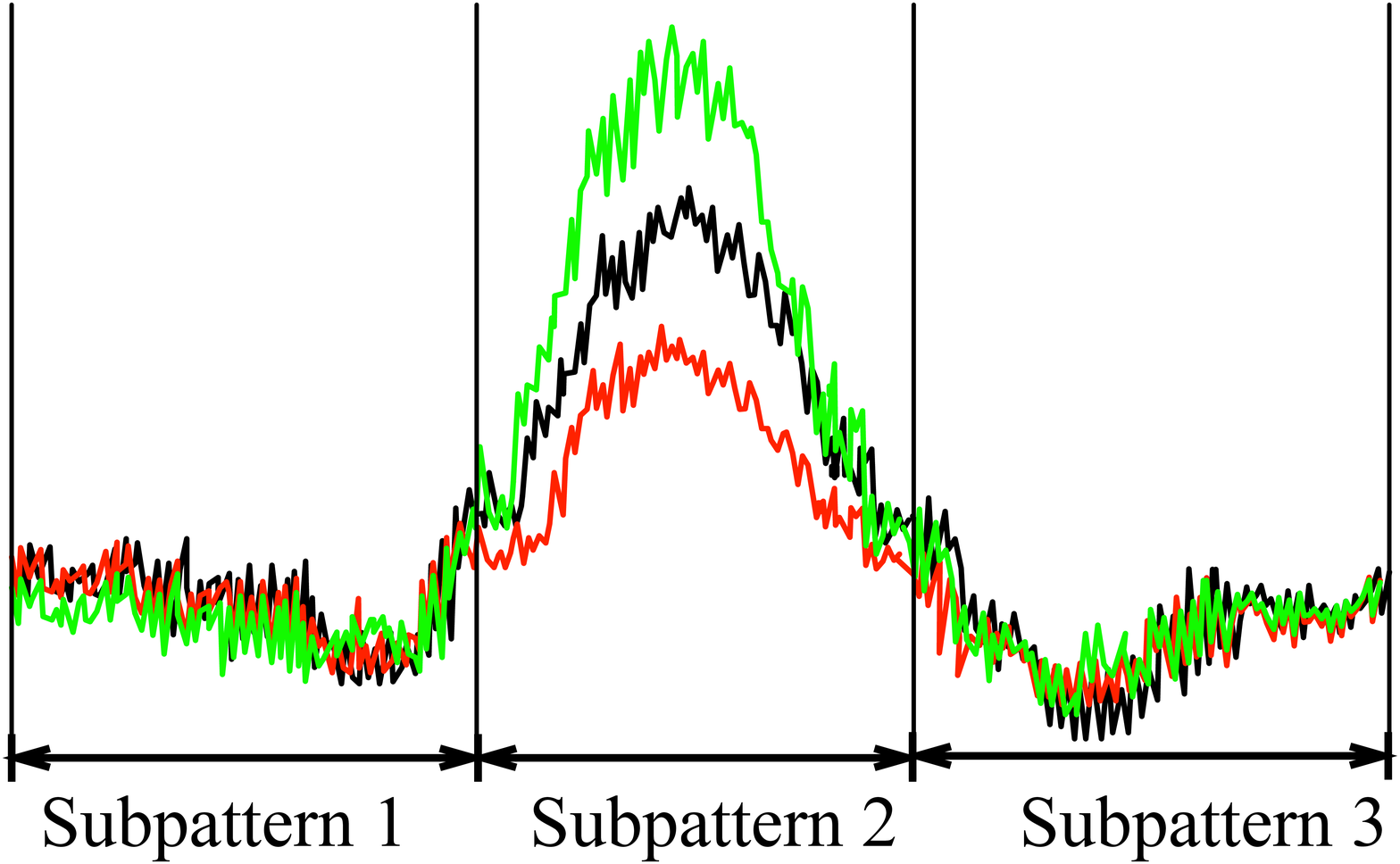}
	\caption{
		Diverse patterns of Extreme Operating Gust(EOG). EOG pattern is composed of three subpatterns and users tend to specify a larger threshold for Subpattern 2 comparing with Subpattern 1 and Subpattern 3. 
	}
	\label{fig:intro_apply}
		\end{minipage}
	\hfill
		\begin{minipage}[t]{0.48\textwidth}
		\centering
		\includegraphics[width=\textwidth]{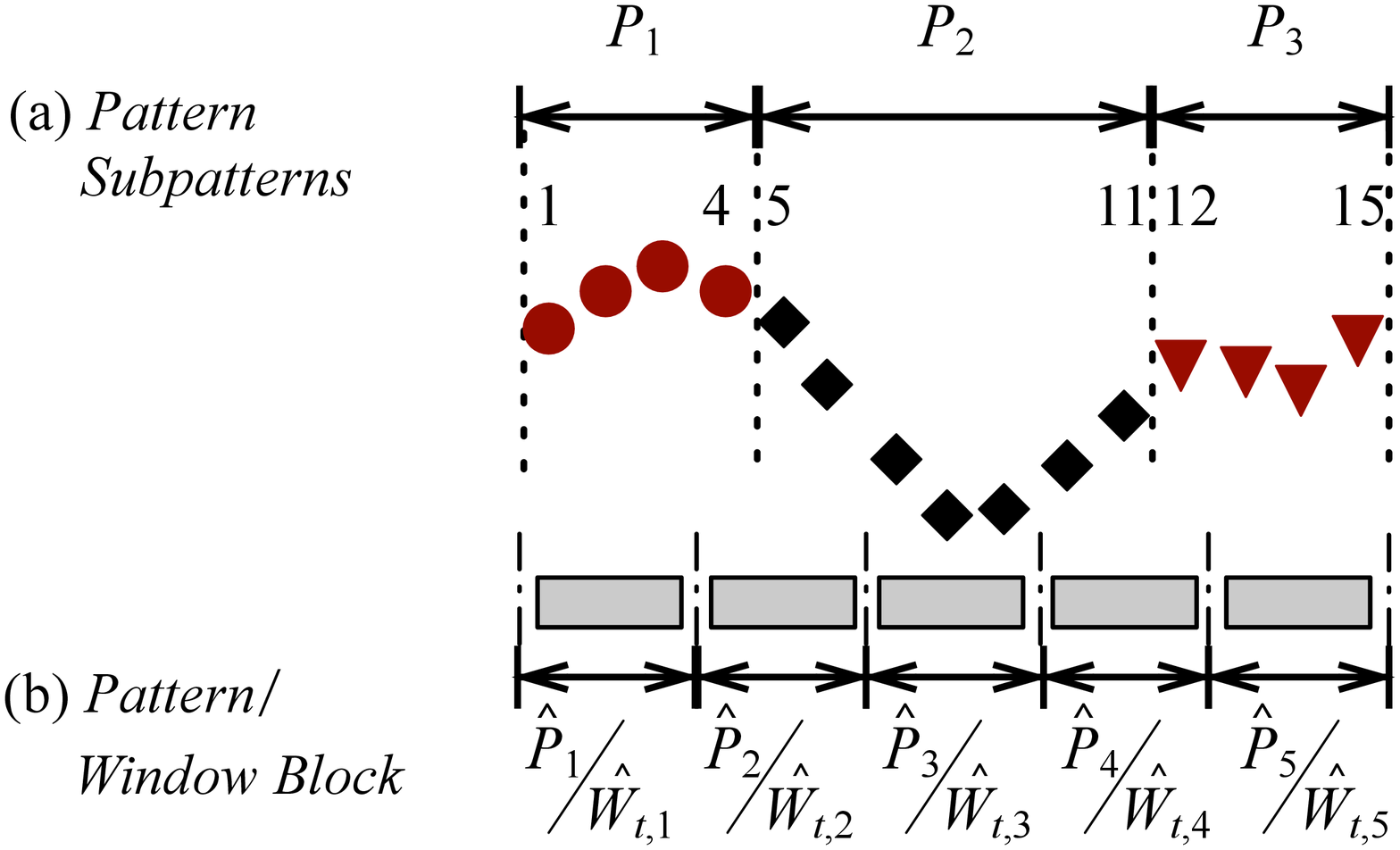}
		\caption{
			(a) Pattern $ P $ is composed of three subpatterns: $ P_1 = P[1:4]$, $ P_2 = P[5:11]$ and $ P_3 = P[12:15]$.\\
			(b) In ELB representation, if we set block size $ w =3 $, $ P $ and $ W_t $ are divided into 5 pattern/window blocks.
		}
	\label{fig:lb_window_block}
	\end{minipage}
\end{figure}

In the field of wind power generation,
Extreme Operating Gust (EOG) \cite{branlard_wind_2009} is a typical gust pattern which is a phenomenon of dramatic changes of wind speed in a short period. Early detection of EOG can prevent the damage to the turbine  \cite{pace_lidar-based_2012}.
A typical pattern of EOG has three physical phases, where its corresponding shape contains a slight decrease (Subpattern 1), followed by a steep rise, a steep drop (Subpattern 2), and a rise back to the original value (Subpattern 3).
Users usually emphasize the shape feature of the second subpattern much more than its exact numeric value. In other words, users tend to specify a larger threshold of distance measurement for Subpattern 2 comparing with Subpattern 1 and Subpattern 3. 
For instance, all time series in Fig.~\ref{fig:intro_apply} are regarded as correct matches of EOG, although they have diverse values in their second subpatterns.

In summary, above example shows that a complex pattern is usually composed of several subpatterns representing different physical meanings, and users may want to specify various thresholds for different parts. 
There are similar situations in other fields like electrocardiogram in Healthcare and technique analysis in the stock market. Therefore, we formulate a new problem, named as \textit{consecutive subpatterns matching} over streaming time series.
In this scenario, a pattern contains a list of consecutive subpatterns with different thresholds. A sliding window on stream matches the given pattern only if each of its components matches the corresponding subpattern.

Although many techniques have been proposed for time series similarity matching, they do not aim to solve the problem mentioned above. For streaming time series matching, some recent works take advantage of similarity or correlation of multiple patterns and avoid the whole matching of every single patterns~\cite{lian_similarity_2007,wei_atomic_2005}. Similarly, most of the previous approaches for subsequence similarity search explore and index the commonalities of time series in database to accelerate the query \cite{loh_subsequence_2004,wang_data-adaptive_2013}. These approaches are not optimized for the scenario of matching consecutive subpatterns.

In this paper, we propose Equal-Length Block (ELB) representation together with the \textit{lower bounding property}. 
ELB representation divides both the pattern and a sliding window into equal-length disjoint pattern/window blocks. Then ELB characterizes a pattern block as upper/lower bounds and a window block as a single value.
Two ELB implementations are provided which allow us to process multiple successive windows together, so that speed up the matching process dramatically while guaranteeing no false dismissals. 

In summary, this paper makes the following contributions:
\begin{itemize}
	\item We introduce a new model, consecutive subpatterns matching, which allows us to describe pattern more expressively and process streaming time series more precisely.
	\item We propose a novel ELB representation which accelerate the matching process dramatically under all $ L_p $-norms and guarantees no false dismissals.
	\item We illustrate the efficiency of our algorithms with sufficient experiments on real-world and synthetic datasets and a comprehensive theoretical analysis.
\end{itemize}

The rest of the paper is arranged as follows:
Sect.~\ref{sec:relate_work} gives a brief review of the related work.
Sect.~\ref{sec:problem_overview} formally defines our problem.
Sect.~\ref{sec:lower_bounding} proposes ELB representation together with  its two implementations.
Sect.~\ref{sec:experiment} conducts extensive experiments.
Finally, Sect.~\ref{sec:conclusion} concludes the paper.

%% file: sec2-related-work.tex
\section{Related Work}
\label{sec:relate_work}
There are two categories of the related works, multiple patterns matching over streaming  time series and subsequence similarity search.

\noindent  \textbf{Multiple patterns matching over streaming time series}.
Traditional single pattern matching over the stream is relatively trivial, hence recent research works put more focus on optimizing the multiple pattern scenario. 
Atomic wedge \cite{wei_atomic_2005} is proposed to monitor stream with a set of pre-defined patterns, which exploits the commonality among patterns.
Sun et al. \cite{sun_matching_2009} extend atomic wedge for various length queries and tolerances.
Lian et al. \cite{lian_similarity_2007} propose a multi-scale segment mean (MSM) representation to detect static patterns over streaming time series.
They discuss the batch processing optimization and the case of dynamic patterns in its following work \cite{lian_multiscale_2009}.
Lim et al. \cite{lim_similar_2008} propose SSM-IS which divides long sequences into smaller windows.
Although these techniques are proposed for streaming time series and some of them speed up the distance calculation between the pattern and the candidate, most of them focus on exploring the commonality and correlation among multiple patterns for pruning unmatched pattern candidates, which doesn't reduce the complexity brought by the problem of consecutive subpatterns matching.

\noindent  \textbf{Subsequence similarity search}.
FRM \cite{faloutsos_fast_1994} is the first work for subsequence similarity search which maps data sequences in database into multidimensional rectangles in feature space.
General Match \cite{moon_general_2002} divides data sequences into generalized sliding windows and the query sequence into generalized disjoint windows, which focuses on estimating parameters to minimize the page access.
Loh et al. \cite{loh_subsequence_2004} propose a subsequence matching algorithm that supports normalization transform.
Lim et al. \cite{lim_using_2006} address this problem by selecting the most appropriate index from multiple indexes built on different windows sizes.
Kotsifakos et al. \cite{kotsifakos_subsequence_2011} propose a framework which allows gaps and variable tolerances in query and candidates.
Wang et al. \cite{wang_data-adaptive_2013} propose DSTree which is a data adaptive and dynamic segmentation index on time series.
This category of researches focuses on indexing the common features of \textit{archived} time series, which is not optimized for pattern matching over the stream.

%% file: sec3-problem-definition.tex
\section{Problem Definition}
\label{sec:problem_overview}

Pattern $P$ is a time series which contains $ n $ number of elements $(p_1, \cdots, p_{n})$. We denote the subsequence $ (p_i, \cdots,  p_j)$ of $P$ by $ P[i:j] $. Logically, $P$ could be divided into several consecutive subpatterns which may have varied thresholds of matching deviation.
Given a pattern $P$, $ P $ is divided into $ b $ number of non-overlapping subsequences in time order, represented as $P_1, P_2,$ $\cdots, P_{b}$, in which the $k$-th subsequence $ P_k $ is defined as the $k $-th subpattern and associated with a specified threshold $ \varepsilon_k $.

As shown in Fig.~\ref{fig:lb_window_block}(a), for instance, pattern $ P $ is composed of three subpatterns: $ P_1 = P[1:4]$, $ P_2 = P[5:11]$ and $ P_3 = P[12:15]$. These subpatterns may be spesified different thresholds.

A streaming time series $S$ is an ordered sequence of elements  that arrive in time order.
We denote a sliding window on $ S $ which starts with timestamp $ t $ by  $ W_t = (s_{t,1}, s_{t,2},\cdots, s_{t, n})$.
We denote the the subsequence $(s_{t,i},s_{t,i+1},\cdots s_{t,j}) $ in $ W_t $ by $ W_t[i:j] $.
According to the sub-pattern division of $ P $,
$ W_t $ is also divided into $ b $ sub-windows $W_{t,1}, W_{t,2},\cdots, W_{t,b}$.
For convenience, we refer to $ p_i $ and $ s_{t,i} $ as an \textit{element pair}.

There are many distance functions such as \textit{DTW} \cite{berndt_using_1994}, \textit{LCSS} \cite{vlachos_discovering_2002} , $L_p$-norm \cite{yi_fast_2000}, etc. 
We choose $L_p$-norm distance which covers a wide range of applications \cite{agrawal_efficient_1993}\cite{faloutsos_fast_1994}\cite{luo_piecewise_2015}.
Given two $ n $-length sequences where $X=(x_1,x_2,\cdots, x_n)$ and $Y=(y_1,y_2,\cdots,y_n)$, the \textit{$L_p$-Norm} Distance between $X$ and $Y$ is defined as follows:
\begin{displaymath}
L_p(X,Y) = (\sum^{n}_{i=1} |{x_i-y_i}|^p)^{\frac{1}{p}}
\end{displaymath}

Since the $ L_{norm} $ is a distance function between two equal-length sequences, there are $ |W_t| = n $ and $ |W_{t,k}| = |P_k| $ for $ k \in [1,b] $.
In addition, we denote by $ L_p[i:j] $ the normalized Euclidean distance  between $ P[i:j] $ and $ W_t[i:j] $.

\noindent \textbf{Problem Statement}:
	Given a pattern $P$ which contains $ b $ number of subpatterns $ P_1, P_2, \cdots, P_b $ with specified thresholds $ \varepsilon_1, \varepsilon_2, \cdots, \varepsilon_b$. 
	For a stream $ S $, \textit{consecutive subpatterns matching}
	is to find all sliding windows $ W_t $ on $ S $, where it holds that $L_p(P_k,W_{t,k}) \leqslant \varepsilon_k$ for $ k \in [1,b] $ (denoted by $ W_{t,k} \prec P_k$).

%% file: sec4-elb.tex
\section{Equal-Length Block}\label{sec:lower_bounding}
In this section, we first sketch a novel representation, Equal-Length Block(ELB), together with \textit{Lower Bounding Property}, which enables us to process several successive windows together while guaranteeing no false dismissals.
After that, we will introduce two ELB implementations in turn. 

ELB representation is inspired by the following observation. 
To avoid false dismissals, a naive method is to slide the window over the stream by one element and calculates the corresponding distance, which is computationally expensive. However, one interesting observation is that in most real-world applications, the majority of adjacent subsequences of time series might be similar. This heuristic gives us the opportunity to process multiple successive windows together. Based on this hint, we propose Equal-Length Block (ELB), and the corresponding lower bounding property. 

ELB divides the pattern $P$ and the sliding window $ W_t $ into several disjoint $ w $-length \textit{block}s while the last indivisible part can be safely discarded. 
The block division is independent of pattern subpatterns. A block may overlap with two or more adjacent subpatterns, and a subpattern may contain more than one block. 
The number of blocks is denoted by $N = \lfloor n/w \rfloor$.
Based on the concept of block, $P$ and $W_t$ are split into $\hat{P} = \{\hat{P}_1, \cdots, \hat{P}_{N}\}$ and $ \hat{W}_t = \{\hat{W}_{t, 1}, \cdots, \hat{W}_{t,N}\} $ respectively, where $\hat{P}_j $ (or $\hat{W}_{t,j}$) is the $j$-th block of $P$ (or $W_t$), that is, $\hat{P}_j =\{p_{(j-1)\cdot w+1},p_{(j-1)\cdot w+2},\cdots, p_{j\cdot w}\} $,
similarly for $ \hat{W}_{t,j} $. 
As shown in Fig.~\ref{fig:lb_window_block}(b),
we set $ w = 3$, thus $P$ and $W_t$ are divided into 5 blocks.
Based on blocks, each pattern block $ \hat{P}_j $ is represented by a pair of bounds, upper and lower bounds, which are denoted by $ \hat{P}^u_j $ and $ \hat{P}^l_j $ respectively. Each window block $ \hat{W}_{t, j} $ is represented by a feature value, denoted by $ \hat{W}_{t, j}^f$. 

It is worth noting that the ELB representation is only an abstract format description, which doesn't specify how to compute upper and lower bounds of $\hat{P}_j$ and the feature of window $ \hat{W}_{t, j}$. 
We can design any ELB implementation, which just needs to satisfy the following lower bounding property:
\begin{definition}\label{def:lower_bounding_property}
	(\textbf{Lower Bounding Property}):
	given $ \hat{P} $ and $ \hat{W}_t $,
	if $\exists \ i \in [0,w)$, $W_{t+i}$ is a result of consecutive subpatterns matching of $P$,
	then $\forall j \in [1,N]$,
	$ \hat{P}^l_j \leqslant \hat{W}_{t, j}^f \leqslant \hat{P}_j^u$ (marked as $\hat{W}_{t, j} \prec \hat{P}_j$).
\end{definition}


We first provide our matching algorithm based on ELB which satisfies lower bounding property before introducing our ELB implementation. Instead of processing sliding windows one-by-one, lower bounding property enables us to process $ w $ successive windows together in the pruning phase.
Given $ N $ number of window blocks $ \{\hat{W}_{t, 1}, \cdots, \hat{W}_{t, N}\} $, if anyone in them (e.g. $\hat{W}_{t,j}$)  doesn't match its aligned pattern block ($\hat{P}_{j}$ correspondingly),  we could skip $w$ consecutive windows, $ W_t, W_{t+1},\cdots, W_{t+w-1} $, together. Otherwise, the algorithm takes these $w$ windows as candidates and calculate exact distances one by one.
The lower bounding property enables us to extend the sliding step to $ w $ while guaranteeing no false dismissals. The critical challenge is how to design ELB implementation which is both computationally efficient and effective to prune sliding windows. 

\subsection{Element-based ELB Representation} 
\label{sub:lb_element}
In this section, we present the first ELB implementation, element-based ELB, denoted by $ELB_{ele}$. 
The basic idea is as follows. According to our problem statement, if window $W_t $ matches $P$, for any subpattern $P_k$ and corresponding $W_{t,k}$, their $ L_p $-Norm distance holds that:
\begin{equation}
L_p(W_{t,k},P_k) \leqslant \varepsilon_k
\label{eq:max_torlerance}
\end{equation}
It's easy to infer that any element pair $p_i$ together with $s_{t,i} $, which falls into the $ k $-th subpattern, satisfies that:
\begin{equation}
|s_{t,i}-p_i| \leqslant \varepsilon_k
\label{eq:rationale}
\end{equation}
In other words, if $s_{t,i}$ falls out of the range
$ [p_i - \varepsilon_k,p_i + \varepsilon_k] $, we know that $W_t$ cannot match $P$.

\begin{figure}[!htb]
	\centering
	\includegraphics[width=0.75	\textwidth]{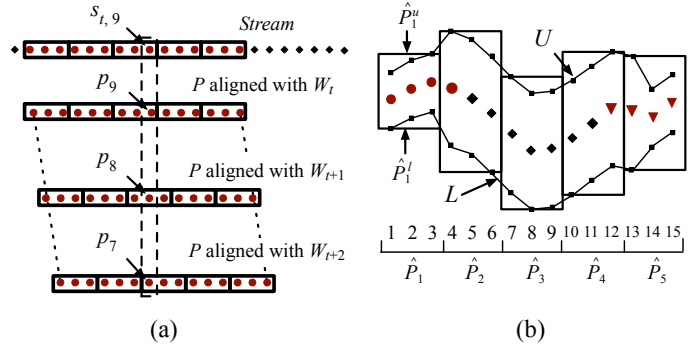}
	\caption{
		(a) The element $ s_{t,9} $ aligns with $ p_9, p_8$ and $p_7 $ at $W_{t},W_{t+1}$ and $W_{t+2}$ respectively.
		(b) $ \hat{P}_j^u $ and $ \hat{P}_j^l $ are constructed by $ U $ and $ L $.
	}
	\label{fig:03elb-ele}
\end{figure}

Based on this observation, we construct two envelope lines for pattern $P$, as illustrated in Fig.~\ref{fig:03elb-ele}(b).
The upper line $ U = \{U_1, U_2, \cdots, U_n\} $ and the lower line $ L = \{L_1, L_2, \cdots, L_n\} $ 
are defined as follows, $ 1 \leqslant i \leqslant n $:
\begin{equation}
\begin{cases}
\begin{split}
U_{i} = p_i+\varepsilon_k \\
L_{i} = p_i-\varepsilon_k
\end{split}
\end{cases}
\label{eq:element_ul}
\end{equation}
The envelope guarantees that if $s_{t,i}$ falls out of $[L_{i},U_i]$, we know that $W_t$ cannot match $P$.

Now we consider how to construct ELB implementation satisfying the  lower bounding property, i.e., how to construct upper/lower bounds of pattern block and the feature of window block so that we could prune $ w $ number of successive windows together. We show the basic idea with an example in Fig.~\ref{fig:03elb-ele}(a). Assume $w=3$ and $ N = 5 $.
At the sliding window $W_{t}$, 
element $s_{t,9}$ aligns with $ p_{9} $. Accordingly, in $W_{t+1}$ (or $W_{t+2}$), $s_{t,9}$ aligns with $p_{8} $ (or $ p_{7} $).
Obviously, if $s_{t,9}$ falls out of all upper and lower envelopes of
$ p_{9} $, $ p_{8} $ and $ p_{7} $, these 3 corresponding windows can be pruned together. 
Note that $s_{t,9}$ is the last element of block $\hat{W}_{t,3}$, and only in this case, all three elements of $P$ aligning with $s_{t,9}$ belong to a same pattern block $\hat{P}_3$.
Based on this observation, we define $ \hat{P}_j^u $, $ \hat{P}_j^l $ and $ \hat{W}_j^f $ as follows:
\begin{equation}
\begin{cases}
\begin{split}
&\hat{P}_{j}^u = \max \limits_{0 \leqslant i < w} (U_{j\cdot w - i})\\
&\hat{P}_{j}^l = \min \limits_{0 \leqslant i < w} (L_{j\cdot w - i})\\
&\hat{W}_{t,j}^f = last(\hat{W}_{t,j}) = s_{t,j\cdot w}
\end{split}
\end{cases}
\label{eq:element_bound}
\end{equation}
As shown in Fig.~\ref{fig:03elb-ele}(b), for each pattern block,
its upper and lower bounds are set to the maximum and minimum of its two envelope lines respectively. It's obvious that $ ELB_{ele} $ satisfies the lower bounding property.

\subsection{Subsequence-based ELB Representation} 
\label{sub:lb_avg}
In this section, we introduce the second ELB implementation, subsequence-based ELB, denoted by $ELB_{seq}$. 
Compared to $ELB_{ele}$,  $ELB_{seq}$ has a tighter bound which brings higher pruning power, although it is a little costlier on computing features of window blocks.

Different from $ELB_{ele}$ which uses the tolerance of the whole subpattern to constrain one element pair,  in $ELB_{seq}$, we use the same tolerance to constrain a $w$-length subsequence.
Referring to \cite{lian_multiscale_2009}, 
given two sequences $X=(x_1,\cdots,x_w)$ and $Y=(y_1,\cdots,y_w)$ ,  it holds that:
\begin{equation}
\label{eq:convex}
w\left| \mu_x-\mu_y \right|^p \leqslant \sum_{i=1}^w \left| x_i-y_i \right|^p
\end{equation}	
where $\mu_x$ and $\mu_y$ are the mean values of $X$ and $Y$.
This theorem allows us to construct upper/lower envelope with the mean value of the subsequence. 

Consider two $ w $-length subsequences $ P[i':i] $ and $ W_t[i':i] $ where $ i' = i-w+1 $ ($ i' > 0$ so $ i \geqslant w $).
We first consider the case that all elements in $ P[i':i] $ (or $ W_t[i':i] $) belongs to only one subpattern(like $ P_k $) and the corresponding subwindow(like $ W_{t,k} $). 
If $W_{t,k}$ matches $ P_k$, referring to Eq. \ref{eq:max_torlerance},
we know that:
\begin{equation}
L_p(P[i':i],W_t[i':i])^p =  \sum_{j=i'}^{i}(p_j-s_{t,j})^p \leqslant \varepsilon_k^p
\label{eq:itoi}
\end{equation}
We denote by $\mu_{P[i':i]}$ and $\mu_{W_t[i':i]}$ that the mean value of $P[i':i]$ and $W_t[i':i]$ respectively.  By combining Eq.~\ref{eq:convex} and Eq.~\ref{eq:itoi}, we have:
\begin{equation}
|\mu_{P[i':i]}-\mu_{W_t[i':i]}| \leqslant (\frac{1}{w} \varepsilon_k^p)^{1/p}
\end{equation}
We construct the envelope of pattern $P$ as follows, $ w \leqslant i \leqslant n $:
\begin{equation}
\begin{cases}
\begin{split}
U_i = \mu_{P[i':i]}+ (\frac{1}{w} \varepsilon_k^p)^{1/p} \\
L_i = \mu_{P[i':i]}- (\frac{1}{w} \varepsilon_k^p)^{1/p}
\end{split}
\end{cases}
\label{eq:onesegment}
\end{equation}

Now we consider the case that the interval $[i':i]$ overlaps with more than one subpattern. Suppose $P[i':i]$ overlaps with $P_{k_l}, P_{k_l+1}, \cdots, P_{k_r}$.
Due to the additivity of the $ p $-th power of $ L_p $-Norm,  we deduce from Eq.~\ref{eq:itoi} that:
\begin{equation}
L_p(P[i':i],W_t[i':i])^p =  \sum_{j=i'}^{i}(p_j-s_{t,j})^p \leqslant \sum_{k=k_l}^{k_r}  \varepsilon_k^p
\label{eq:dedue-itoi}
\end{equation}
By combining Eq.~\ref{eq:convex} and Eq.~\ref{eq:dedue-itoi}, we have that:
\begin{equation}
|\mu_{P[i':i]}-\mu_{W_t[i':i]}|
\leqslant (\frac{1}{w}\sum_{k=k_l}^{k_r} \varepsilon_k^p)^{1/p}
\label{eq:seq_limit}
\end{equation}
We denoted the right term as $\theta_{seq}(i)$ and provide the general case of the pattern envelope as follows, $ w \leqslant i \leqslant n $:
\begin{equation}
\begin{cases}
\begin{split}
U_i = \mu_{P[i':i]}+\theta_{seq}(i) \\
L_i = \mu_{P[i':i]}-\theta_{seq}(i)
\end{split}
\end{cases}
\label{eq:avg_ul}
\end{equation}
Note that Eq.~\ref{eq:onesegment} is the special case of Eq.~\ref{eq:avg_ul}.

\begin{figure}[!htb]
	\centering
	\includegraphics[width=0.75\textwidth]{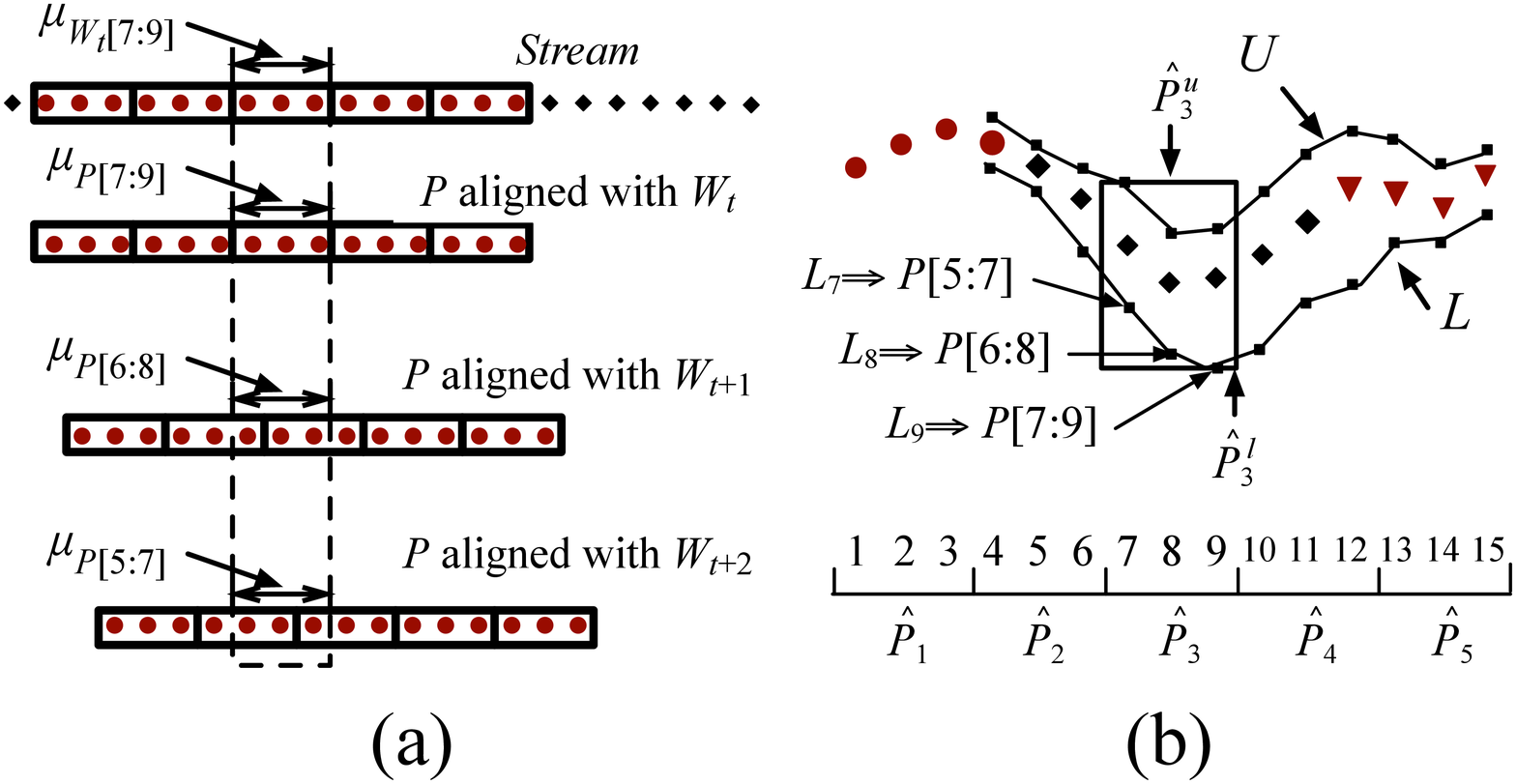}
	\caption{(a) The subsequence $ W_t[7 : 9] $ aligns with $ P[7:9],P[6:8]$ and $P[5:7] $ at $W_{t},W_{t+1}$ and $W_{t+2}$ respectively.
		(b) $ \hat{P}_j^u $ and $ \hat{P}_j^l $ are constructed by $ U $ and $ L $.
	}
	\label{fig:03elb-seq}
\end{figure}

The construction of upper and lower bounds are very similar to $ELB_{ele}$, while the feature of window block is adopted to the mean value. 
We show the basic idea with an example in Fig.~\ref{fig:03elb-seq}(a). At the sliding window $W_{t}$, the subsequence $ W_t[7 : 9] $ aligns with $ P[7:9] $. Similarly, in $W_{t+1}$ (or $W_{t+2}$), this subsequence aligns with $ P[6:8] $ (or $ P[5:7] $).
According to Eq.~\ref{eq:avg_ul}, we know that if the mean value of $W_t[7 : 9] $ falls out of all upper and lower bounds of $ P[7:9], P[6:8] $ and $ P[5:7] $, these 3 corresponding windows can be pruned together. Based on this observation, we give the formal implementation of $ELB_{seq}$ as follows:
\begin{equation}
\begin{cases}
\begin{split}
&\hat{P}_{j}^u = \max \limits_{0 \leqslant i < w} (U_{j\cdot w - i})\\
&\hat{P}_j^l = \min \limits_{0 \leqslant i < w} (L_{j\cdot w - i})\\
&\hat{W}_{t,j}^f = mean(\hat{W}_{t,j}) = \mu_{W_{t}[(j-1)\cdot w +1\, :\, j\cdot w]}
\end{split}
\end{cases}
\label{eq:avg_bound}
\end{equation}
Note that, the upper and lower bounds of $ \hat{P}_1 $ are meaningless according to the definition of the envelope of $ ELB_{seq} $.

Figure~\ref{fig:03elb-seq}(b) provides an example of $ ELB_{seq} $ implementation. For clarity, we only illustrate the bounds of $ \hat{P}_3 $. 
The lower bound $ \hat{P}^l_3 $ is set to the minimum of $ L_7, L_8 $ and $ L_9 $ and covers 3 successive windows $ W_{t}, W_{t+1} $ and $ W_{t+2} $.

\subsection{Complexity Analysis} 
\label{sub:elb_complexity}
We first analyze $ ELB_{ele} $.
For each block $\hat{W}_{t,j} $, the time complexities of  computing feature and determining $\hat{W}_{t,j} \prec \hat{P}_j$ are both $ O(1) $.
Therefore, the amortized pruning cost of $ ELB_{ele} $ is $ O(1/w) $.
Its space complexity is $ O(N) = O(\lfloor n/w \rfloor)$.
Although $ELB_{ele}$ is very efficient, it constrains one element pair with the tolerance of the whole subpattern, which makes the envelope loose.
Its pruning effectiveness is better when thresholds are relatively small, or pattern deviates from the normal stream far enough. 

$ ELB_{seq} $ calculates the mean value of each window block with $ O(w) $ and determining $\hat{W}_{t,j} \prec \hat{P}_j$ with $ O(1) $.
Considering a window block appears in several consecutive sliding windows, we store feature values in memory to avoid repeated calculation.
Therefore, the amortized pruning cost of $ ELB_{seq} $ is reduced to $ O(1) $.
Same as $ ELB_{ele} $, the space complexity of $ ELB_{seq} $ is $ O(N)$.

%% file: sec5-experiment.tex
\section{Experimental Evaluation}
\label{sec:experiment}
In this section, we first describe datasets and experimental settings in Sect.~\ref{sub:experimental_Settings} and then
present the results of performance evaluation comparing the brute-force approach Sequential Scanning(SS), the classic method MSM \cite{lian_multiscale_2009} and our two approaches based on $ ELB_{ele} $(ELB-ELE) and $ ELB_{seq} $ (ELB-SEQ) respectively. 
As presented in Sect.~\ref{sec:relate_work}, although there are many works after MSM addressing time series similarity matching, most of them focus on utilizing the commonality among multiple patterns to build indexes, but not speeding up the problem of matching stream with a list of consecutive subpatterns.

Our goal is to:
\begin{itemize}
	\item Demonstrate the efficiency of our approach on all $ L_p $-Norm distance and different thresholds.
	\item Demonstrate the robustness of our approach on different pattern occurrence probabilities. 
	\item Investigate the impact of block size on performance which helps to choose the appropriate parameter.
\end{itemize}

\subsection{Experimental Setup}
\label{sub:experimental_Settings}
The experiments are conducted on both synthetic and real-world datasets.

\noindent \textbf{Datasets}.
Real-world datasets are collected from a wind turbine manufacturer, where each wind turbine has hundreds of sensors generating streaming time series with sampling rate from 20 ms to 7s.
Our experimental datasets are from 3 turbines.
In each turbine, we collect data of 5 sensors including wind speed, wind deviation, wind direction, generator speed and converter power. 
We replay the data as streams with total lengths of $10^8$.
For each stream, a pattern containing consecutive subpatterns with thresholds is given by domain experts.

Synthetic datasets are constructed based on UCR Archive \cite{keogh_welcome_nodate}.
UCR Archive is a popular time series repository, which includes a set of datasets widely used in time series mining researches \cite{begum_rare_2014,keogh_exact_2002,lian_multiscale_2009}. 
To simulate patterns with various lengths, we select four datasets, Strawberry (Straw for short), Meat, NonInvasiveFatalECG\_Thorax1 (ECG for short) and MALLAT whose time series lengths are 235, 448, 750 and 1024. 
Referring to \cite{lian_multiscale_2009}, for each selected UCR dataset, we choose the first time series of class 1 as the pattern and divide it into several subpatterns according to its shape and trend.  Numbers of subpatterns of these four datasets are 5, 6, 8 and 7 respectively.

Concerning threshold of synthetic datasets, 
we define \textit{threshold\_ratio} as the ratio of the average threshold to the value range of this subpattern. Given a \textit{threshold\_ratio} and a subpattern $ P_k $, the $ L_p $-Norm threshold of $ P_k $ is defined by:
\[
\varepsilon_k = |P_k|^{1/p} \times threshold\_ratio \times value\_range(P_k)
\]
In practice, we observe that \textit{threshold\_ratio} being larger than $ 30\% $ indicates that the average deviation from a stream element to its aligned pattern element is more than $ 30\% $ of its value range.
In this case, the candidate may be quite different from given pattern where similarity matching becomes meaningless. 
Therefore, we vary \textit{threshold\_ratio} from 5\% to 30\% 
in Sect.~\ref{sub:performance_of_different_thres_factor}.

As for streaming data of synthetic datasets, referring to \cite{begum_rare_2014}, we first generate a random walk time series $S$ with length of $ 10^8 $ for each UCR dataset.
Element $ s_i $ of $ S $ is $s_i = R + \sum_{j=1}^{i} (\mu_j - 0.5)$, where $\mu_j$ is a uniform random number in $ [0, 1]$.
As value ranges of the four patterns are about -3 to 3, we set $R$ as the mean value $ 0 $.
Then we randomly embed some time series of class 1 of each UCR dataset into corresponding steaming data with certain occurrence probabilities.

\noindent \textbf{Algorithm}.
We compare our approaches to SS and MSM \cite{lian_multiscale_2009}.
SS matches the sliding window one by one. For each window, SS calculates the $ L_p $-Norm distances between all subpatterns and subwindows sequentially.
In our scene, we let MSM build hierarchical grid index for each subpattern.
For fair comparison, we adopt its batch version where the batch size is equal to ELB block size.
We perform three schemes of MSM to choose the best one: stop the pruning phase at the first level of grid index(MSM-1), the second level (MSM-2), or never early stop the pruning phase(MSM-MAX).

\noindent \textbf{Default Parameter Settings}.
There are three parameters for datasets: distance function, threshold and pattern occurrence probability. There is a parameter for our algorithm: block size.
The default distance function is set to $ L_2 $-Norm (i.e., Euclidean distance).
The default value of \textit{threshold\_ratio} and pattern occurrence probability are set to $ 20\% $ and $ 10^{-4} $ respectively.
We set the default value of block size to $ 5\% $ of the pattern length.
The impact of all above parameters will be investigated in following sections.

\noindent \textbf{Performance Measurement}.
We regard the brute-force method SS as the baseline and measure the speedup of MSM and our algorithms. 
Streams and patterns are loaded into memory in advance where data loading time is excluded.
To avoid the inaccuracy due to cold start and random noise, we run all algorithms over 10,000 ms and average them by their cycle numbers.
All experiments are run on 4.00 GHz Intel(R) Core(TM) i7-4790K CPU, with 8GB physical memory.

\subsection{Performance Analysis}
\label{sub:performance of_different_distances}
In this set of experiments, we first show our algorithms together outperform compared approaches on both synthetic and real-world datasets under different $ L_p $-Norm functions and provide detailed analysis.
After that, we perform experiments on diverse synthetic datasets by varying threshold ratio and pattern occurrence probability to demonstrate efficiency and robustness of our approaches. 
At last, we also evaluate the impact of block size for optimal parameter determination.

\subsubsection{Performance under Different $ L_p $-Norm Distance}
\label{sub:performance_of_different_distance}

In this section, we report experiments of ELB-ELE and ELB-SEQ comparing to SS and MSM under different distance functions. We performed these experiments on all real-world and synthetic datasets using $ L_p $-Norm where $p = 1, 2, 3, \infty$. 

\begin{figure*}[!tp]
	\centering
	\begin{subfigure}[b]{0.24\textwidth}
		\centering
		\includegraphics[width=1.05\textwidth]{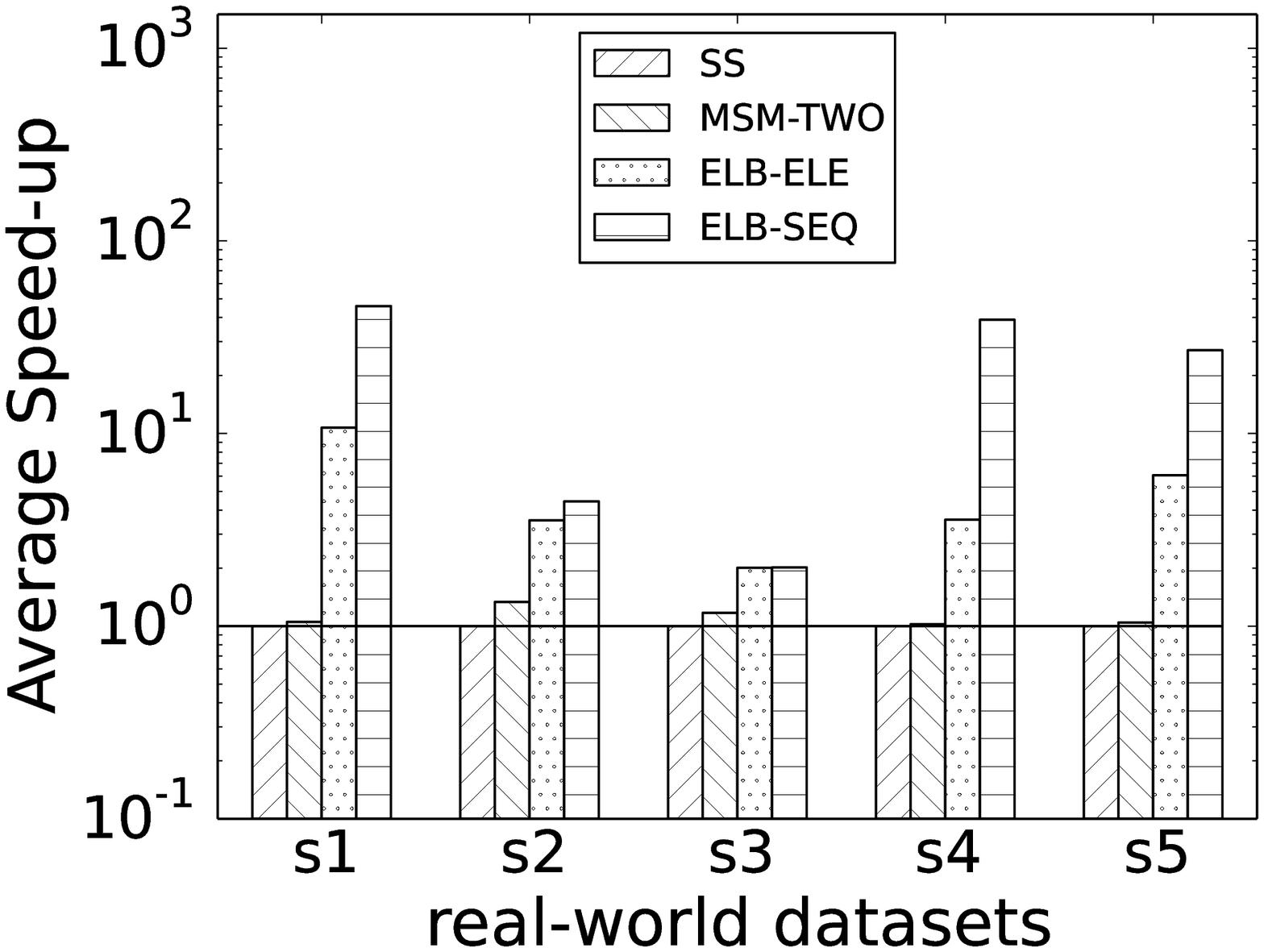}
		\caption[Network2]%
		{wind $ L_1 $-Norm}    
	\end{subfigure}
		\hfill
	\begin{subfigure}[b]{0.24\textwidth}  
		\centering 
		\includegraphics[width=1.05\textwidth]{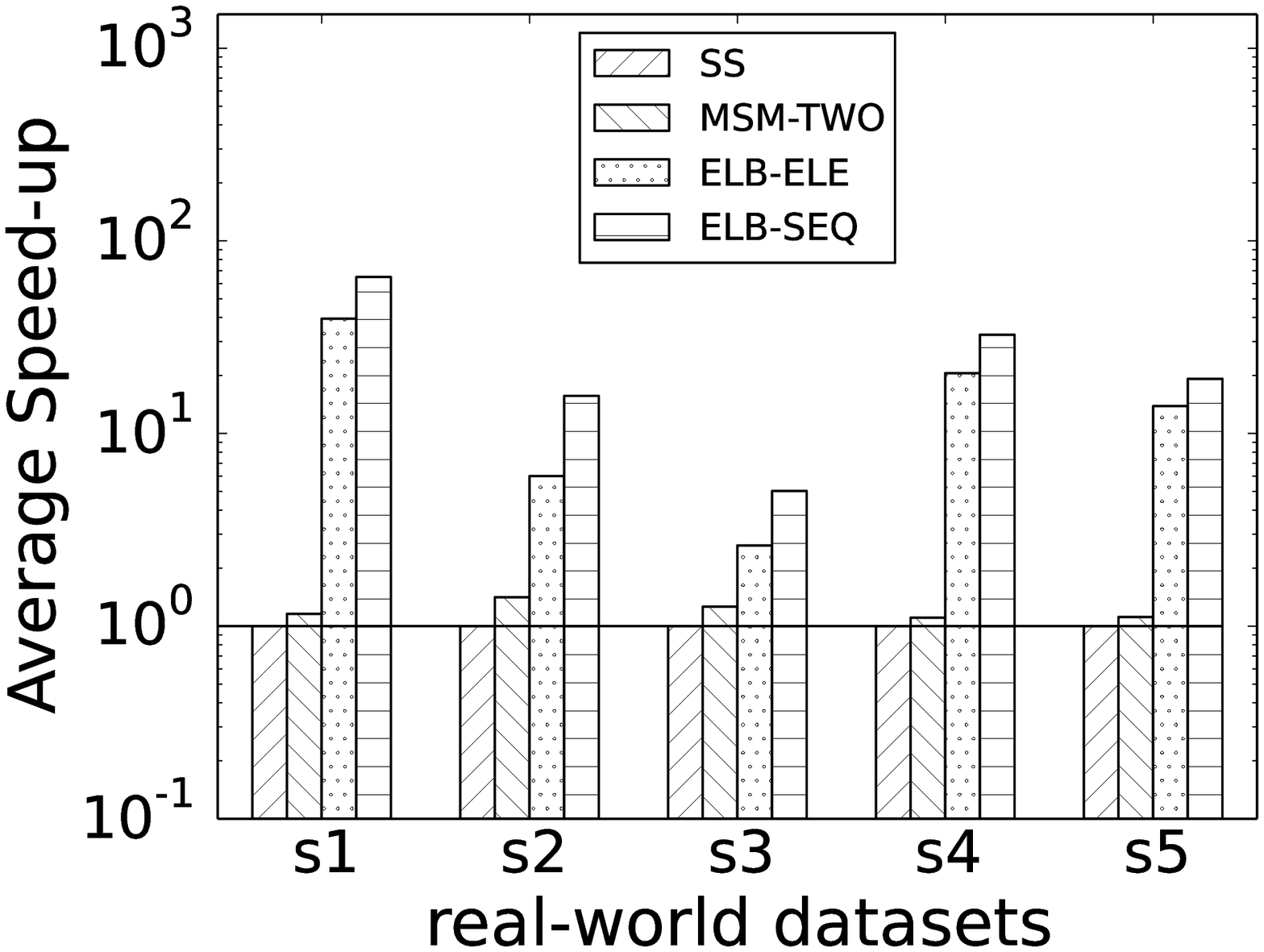}
		\caption[]%
		{wind $ L_2 $-Norm}    
	\end{subfigure}
	\hfill
	\begin{subfigure}[b]{0.24\textwidth}   
		\centering 
		\includegraphics[width=1.05\textwidth]{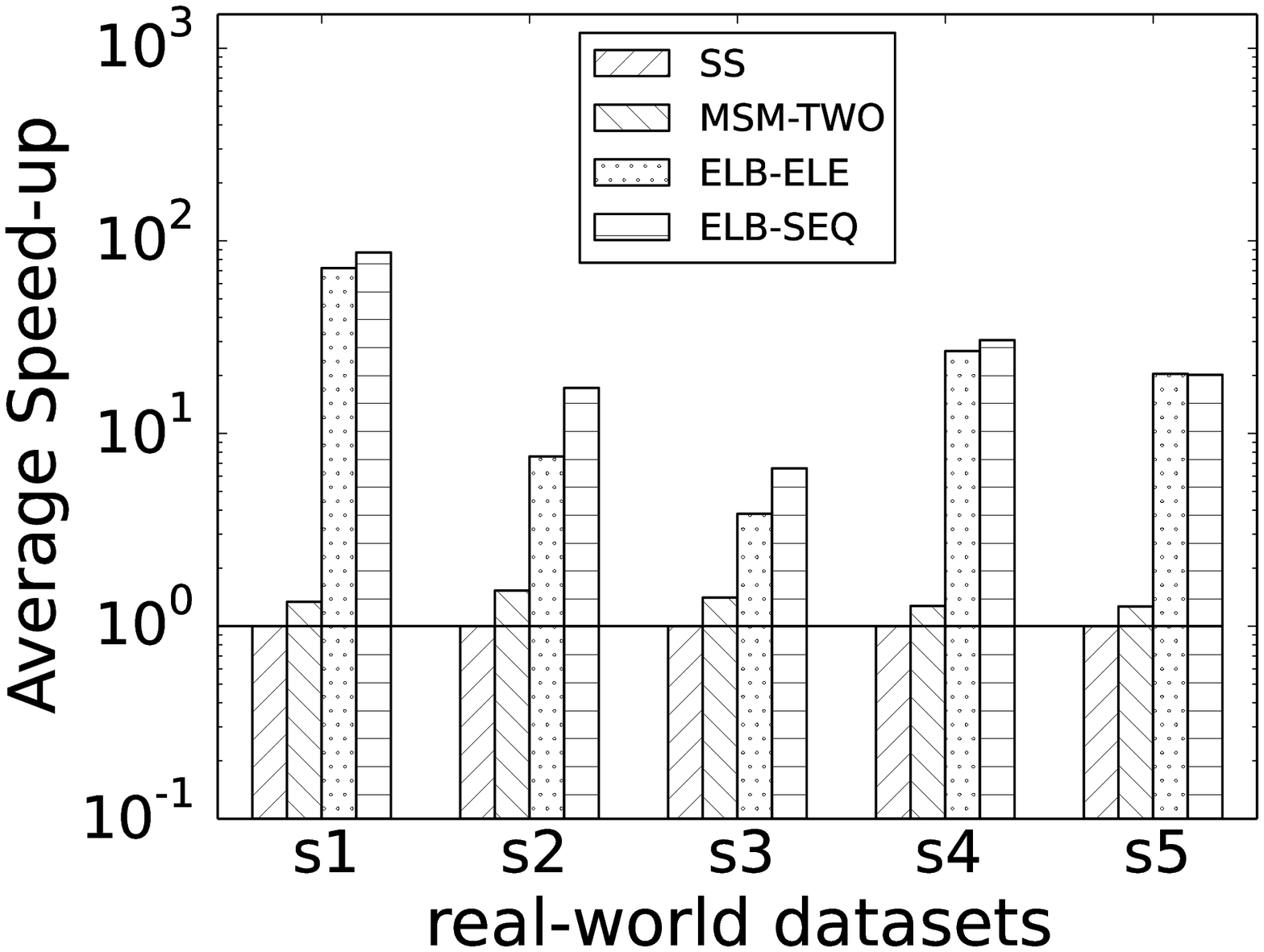}
		\caption[]%
		{wind $ L_3 $-Norm}    
	\end{subfigure}
	\hfill
	\begin{subfigure}[b]{0.24\textwidth}   
		\centering 
		\includegraphics[width=1.05\textwidth]{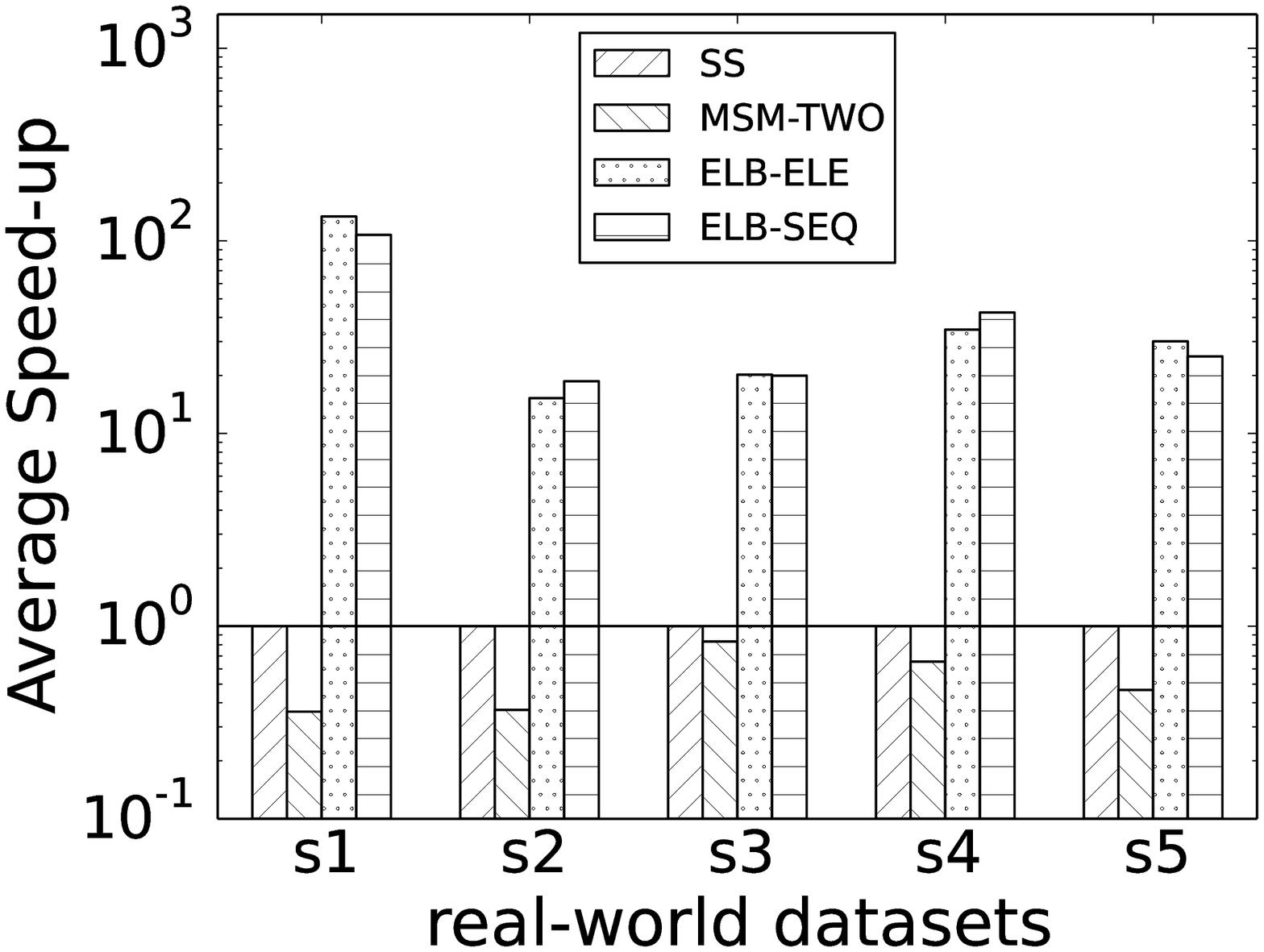}
		\caption[]%
		{wind $ L_\infty $-Norm}    
	\end{subfigure}
	\hfill
	\begin{subfigure}[b]{0.24\textwidth}
		\centering
		\includegraphics[width=1.05\textwidth]{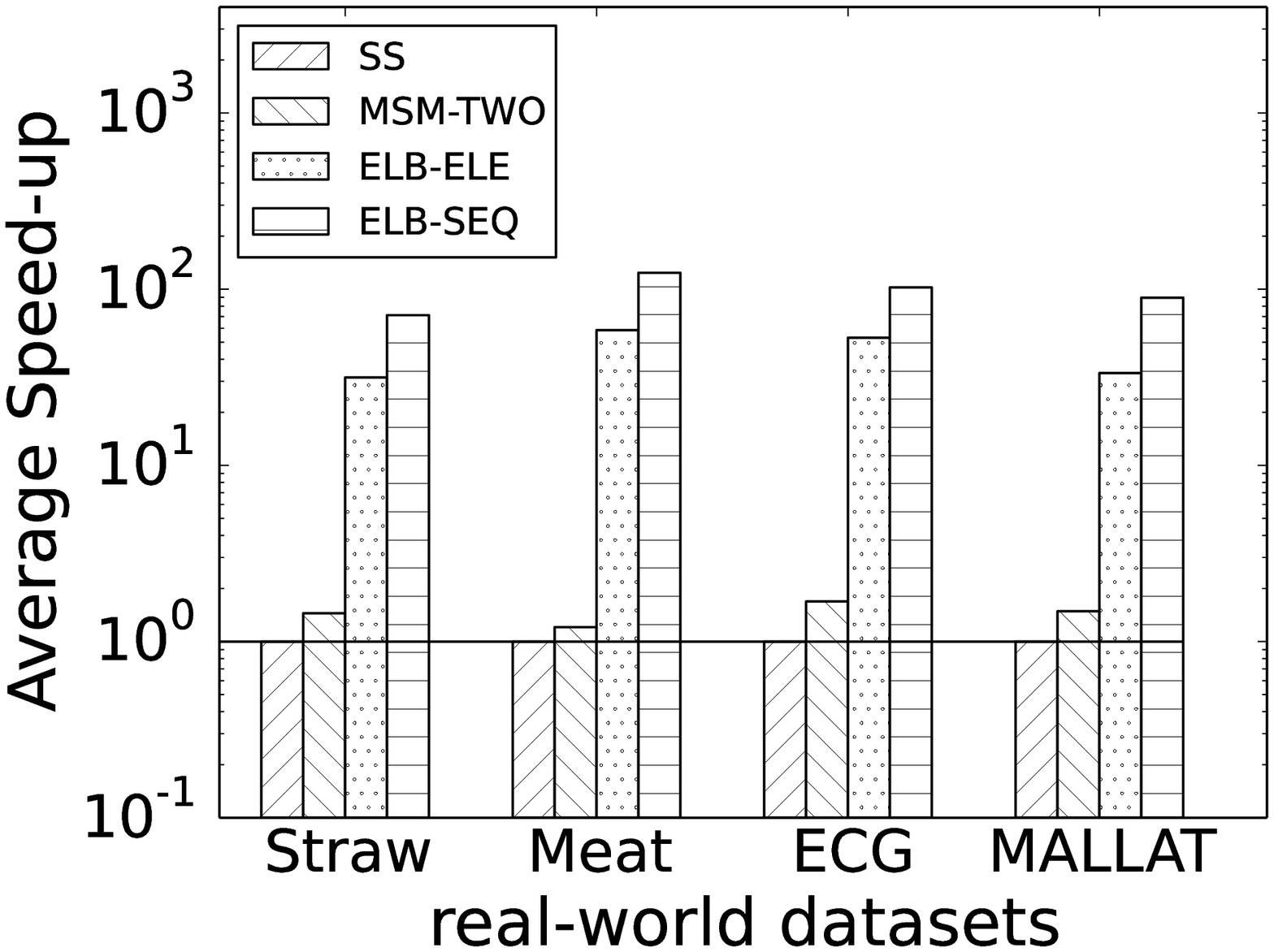}
		\caption[Network2]%
		{UCR $ L_1 $-Norm}    
	\end{subfigure}
	\hfill
	\begin{subfigure}[b]{0.24\textwidth}  
		\centering 
		\includegraphics[width=1.05\textwidth]{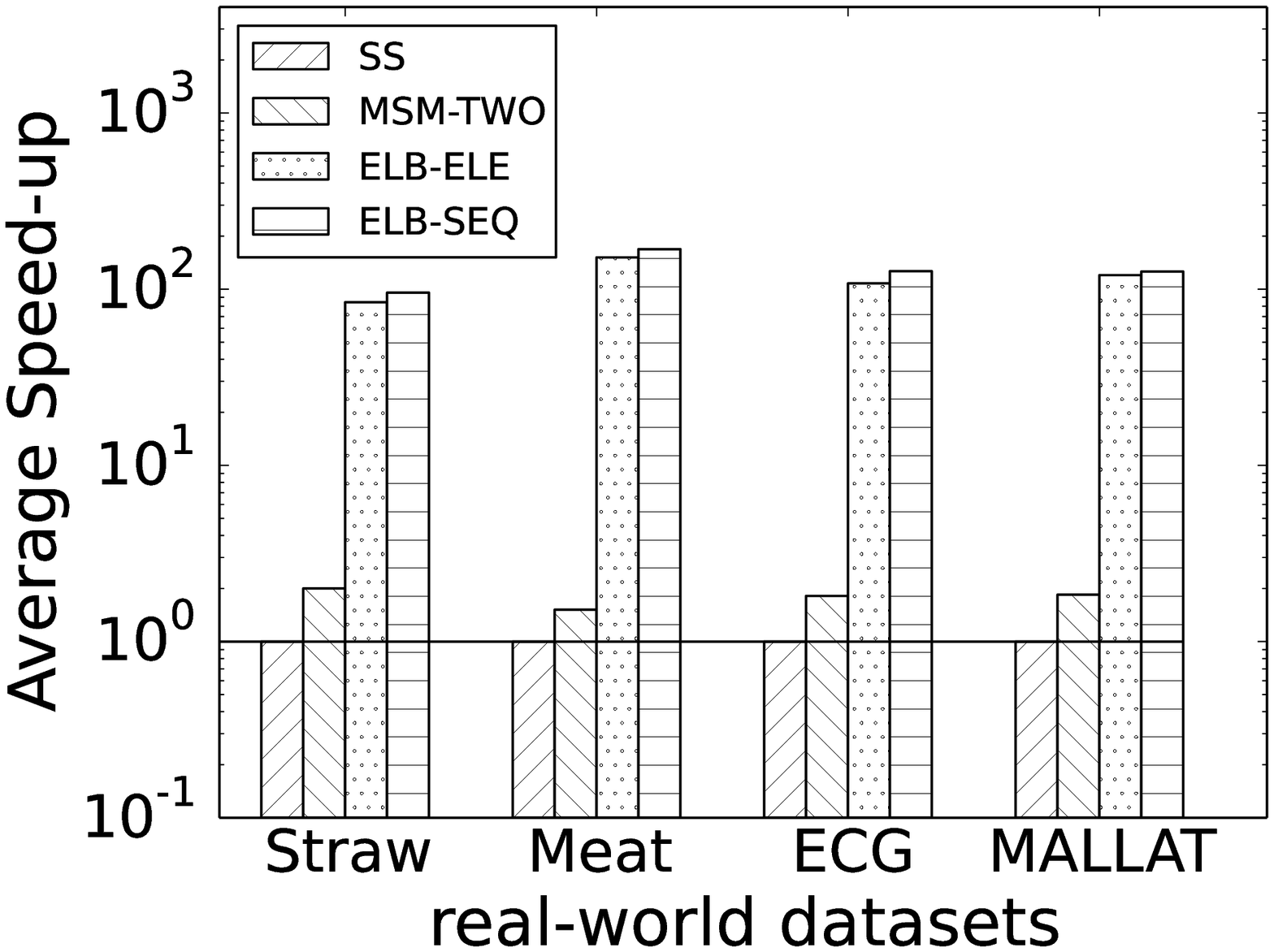}
		\caption[]%
		{UCR $ L_2 $-Norm}    
	\end{subfigure}
	\hfill
	\begin{subfigure}[b]{0.24\textwidth}   
		\centering 
		\includegraphics[width=1.05\textwidth]{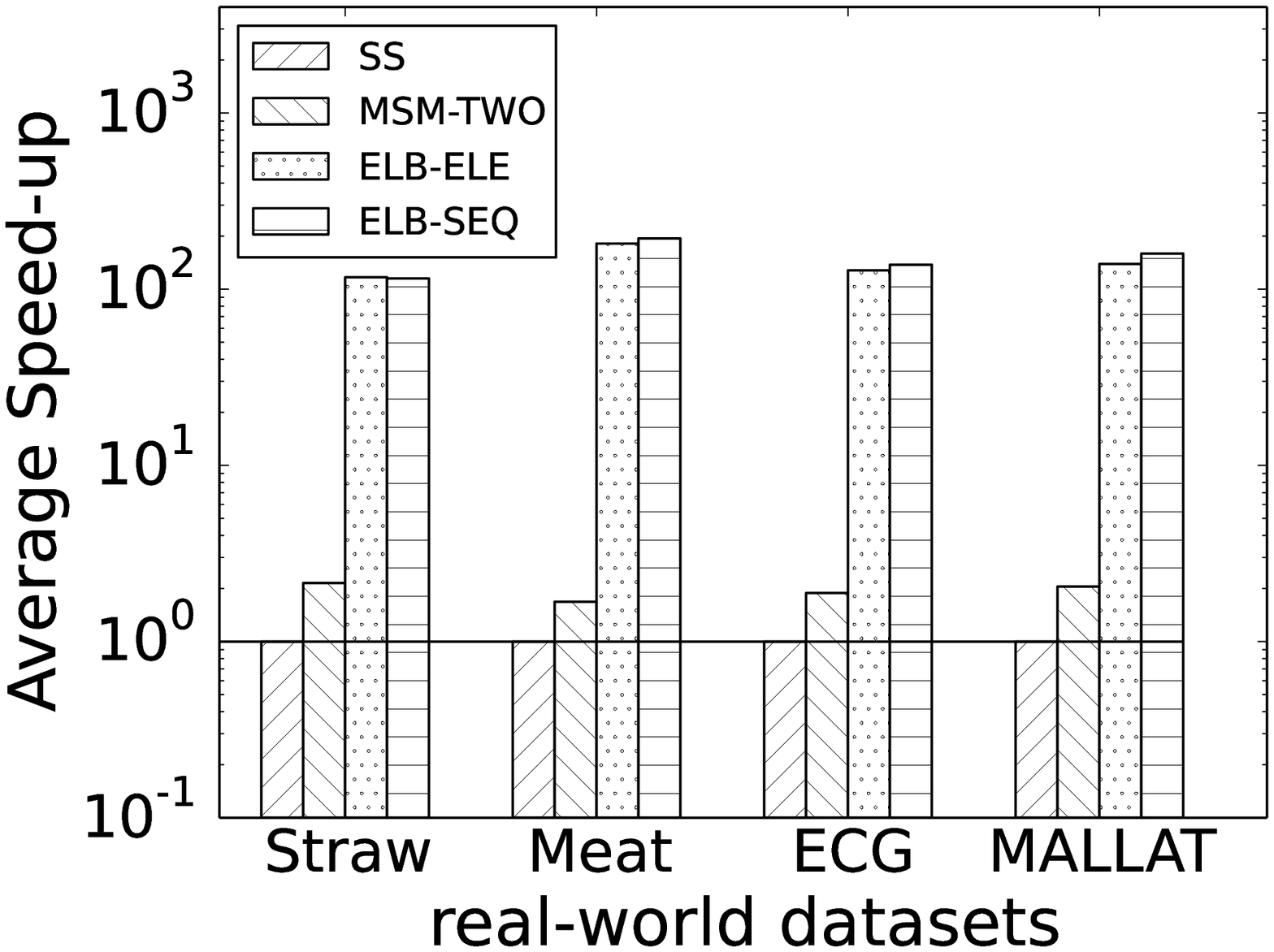}
		\caption[]%
		{UCR $ L_3 $-Norm}    
	\end{subfigure}
	\hfill
	\begin{subfigure}[b]{0.24\textwidth}   
		\centering 
		\includegraphics[width=1.05\textwidth]{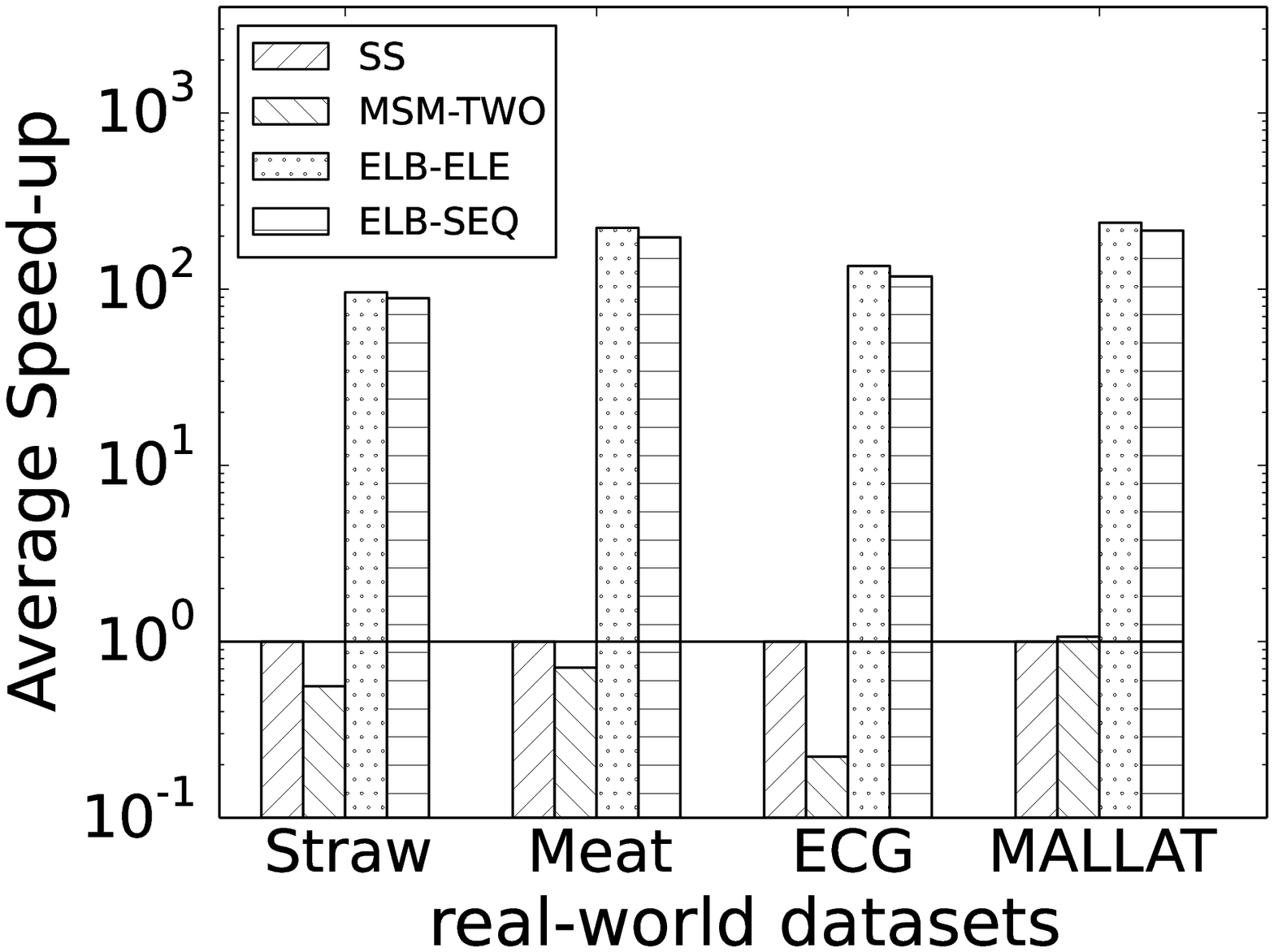}
		\caption[]%
		{UCR $ L_\infty $-Norm}    
	\end{subfigure}
	\caption{Speedup vs. $L_p$-Norm.  
		$ s_1 $:  wind speed, 
		$ s_2 $:  wind deviation, 
		$ s_3 $:  wind direction,
		$ s_4 $:  generator speed,
		$ s_5 $:  converter power.
	}
	\label{fig:5_distance}
\end{figure*}

Figure~\ref{fig:5_distance} shows the experimental results.
For real-world datasets, the results are similar among different turbines, so we only illustrate the wind turbine 1.
Our algorithms show a great advantage over MSM and SS.
As the distance function varies from $L_1$-Norm to $L_\infty$-Norm,
the advantage of our approaches over other methods gets larger.

We provide the experimental detail on a wind generator dataset in Table~\ref{table:distance}.
The first two columns present the total and pruning time on each sliding window.
Column \textit{pruning power} is the percentage of pruned windows.
Comparing to SS, our algorithms could prune numerous windows in the pruning phase, while SS has to perform exact matching for each sliding window, resulting in high time cost.
Regarding MSM, its pruning power gets better from MSM-ONE to MSM-MAX (increased from 98.20\% to 99.96\% in $ L_1 $-Norm).
Although MSM is more accurate, our pruning phase is much more efficient than MSM.
Concerning ELB\_SEQ and MSM\_TWO (the best one among three MSM schemes) on $ L_1 $-Norm, 
our approach has slightly lower pruning power (97.16\% vs. 99.89\%), yet much more efficient pruning cost(0.52 vs. 547.40).
On the whole, ELE\_SEQ has an advantage of more than one order of magnitude over MSM\_TWO.

Now we analyze the different performance of ELB on different $ L_p $-Norm. 
From $ L_1 $-Norm to $ L_\infty $-Norm, the pruning effectiveness of ELB gets better. 
Although ELB-ELE spends less time on pruning phase than ELB-SEQ, its pruning power is very low at $ L_1 $-Norm (6.48\%) due to its too loose bound. 
As $ p $ increases, its bound becomes tighter and performance gets better. In the case of $ L_\infty$-Norm, its performance has been flat with, and even outperformed ELB-SEQ on several datasets, as shown in Fig.~\ref{fig:5_distance}(d) and (h).
In contrast to ELB-ELE, ELB-SEQ is efficient under all $ L_p $-Norms.

\begin{table}[!tp]
	\centering
	\caption{The Detail statistics on wind generator dataset}
	\label{table:distance}
	\begin{tabular}{l|l|l|l|l|l|l}
	\hline
	\multirow{2}{*}{Algorithm} & 
	\multicolumn{3}{c|}{$ L_1 $-Norm}& 
	\multicolumn{3}{c}{$ L_\infty$-Norm}\\ 
	\cline{2-7} 
	& \begin{tabular}[c]{@{}l@{}}total \\ time(ns)\end{tabular} & \begin{tabular}[c]{@{}l@{}}pruning\\ time(ns)\end{tabular} & \begin{tabular}[c]{@{}l@{}}pruning\\ power(\%)\end{tabular} & \begin{tabular}[c]{@{}l@{}}total\\ time(ns)\end{tabular} & \begin{tabular}[c]{@{}l@{}}pruning\\ time(ns)\end{tabular} & \begin{tabular}[c]{@{}l@{}}pruning\\ power(\%)\end{tabular} \\ \hline
ELB\_SEQ                   & \textbf{13.04}                                                 & 0.52                                                   & 97.16                                                    & \textbf{8.78}                                                 & 0.41                                                       & 97.43                                                    \\ \hline
	ELB\_ELE                   & 146.10                                                & 0.51                                                   & 6.48                                                   & 10.78                                                & 0.30                                                       & 96.18                                                    \\ \hline
	MSM\_ONE                   & 556.39                                                & 543.15                                                 & 98.20                                                    & 691.11                                               & 667.34                                                     & 87.05                                                   \\ \hline
	MSM\_TWO                   & 548.42                                                & 547.40                                                 & 99.89                                                    & 670.51                                               & 668.73                                                     & 99.21                                                    \\ \hline
	MSM\_MAX                   & 549.43                                                & 548.94                                                 & 99.96                                                   & 682.40                                               & 682.00                                                     & 99.97                                                    \\ \hline
		SS                         & 562.84                                                & -                                                      & -                                                       & 413.34                                               & -                                                          & -                                                       \\ \hline
	
\end{tabular}

\end{table}

\subsubsection{Impact of Distance Threshold}
\label{sub:performance_of_different_thres_factor}
In this section, we compare the performance of ELB-ELE, ELB-SEQ, SS and MSM under different thresholds.
We vary \textit{threshold\_ratio} from 5\% to 30\% on synthetic datasets,  as described in Sect.~\ref{sub:experimental_Settings}.

\begin{figure}[!tp]
	\centering
	\begin{subfigure}[b]{0.24\textwidth}
		\centering
		\includegraphics[width=1.05\textwidth]{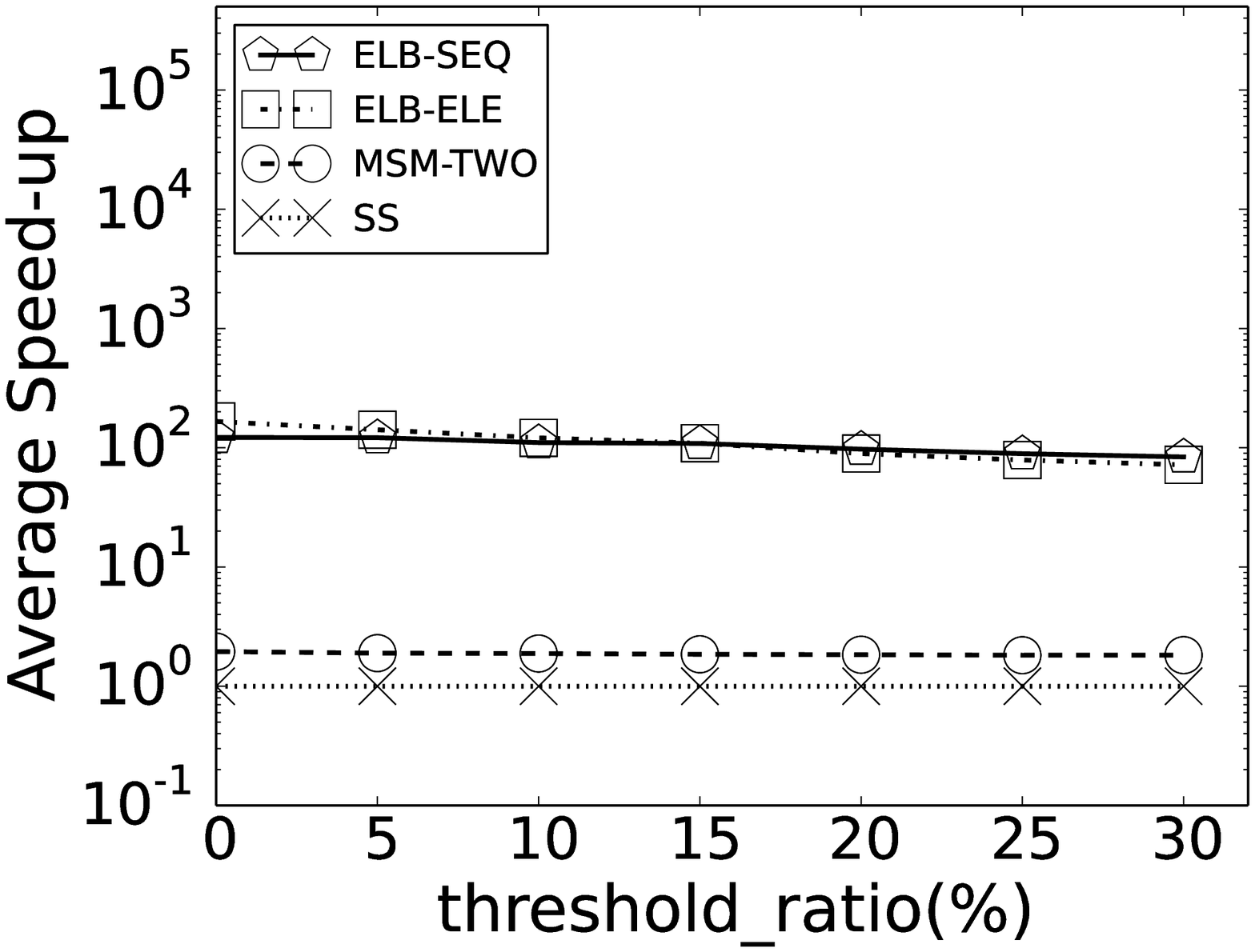}
		\caption[]
		{\small UCR\_Straw}
	\end{subfigure}
	\hfill
	\begin{subfigure}[b]{0.24\textwidth}  
		\centering 
		\includegraphics[width=1.05\textwidth]{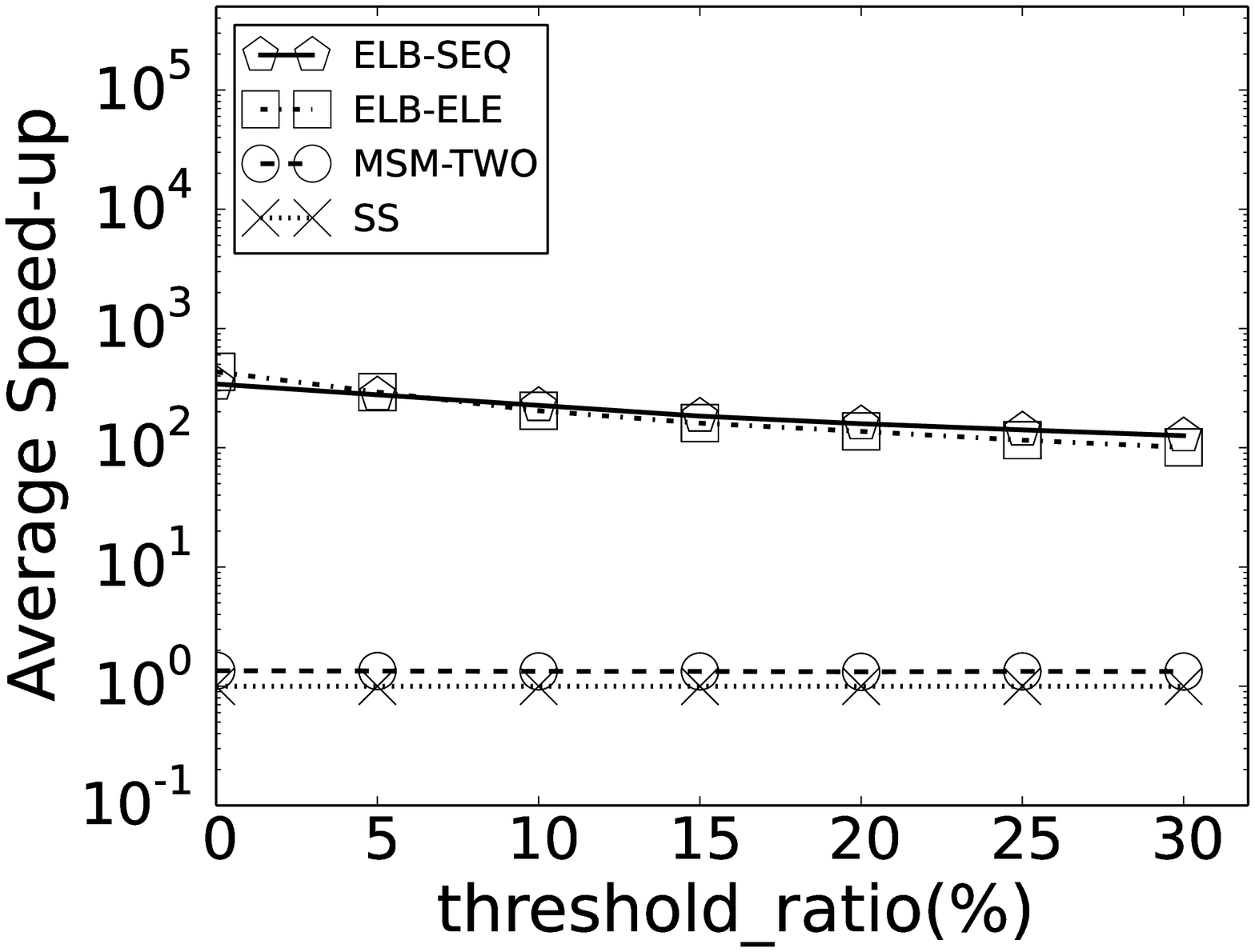}
		\caption[]%
		{UCR\_Meat}    
	\end{subfigure}
	\hfill
	\begin{subfigure}[b]{0.24\textwidth}   
		\centering 
		\includegraphics[width=1.05\textwidth]{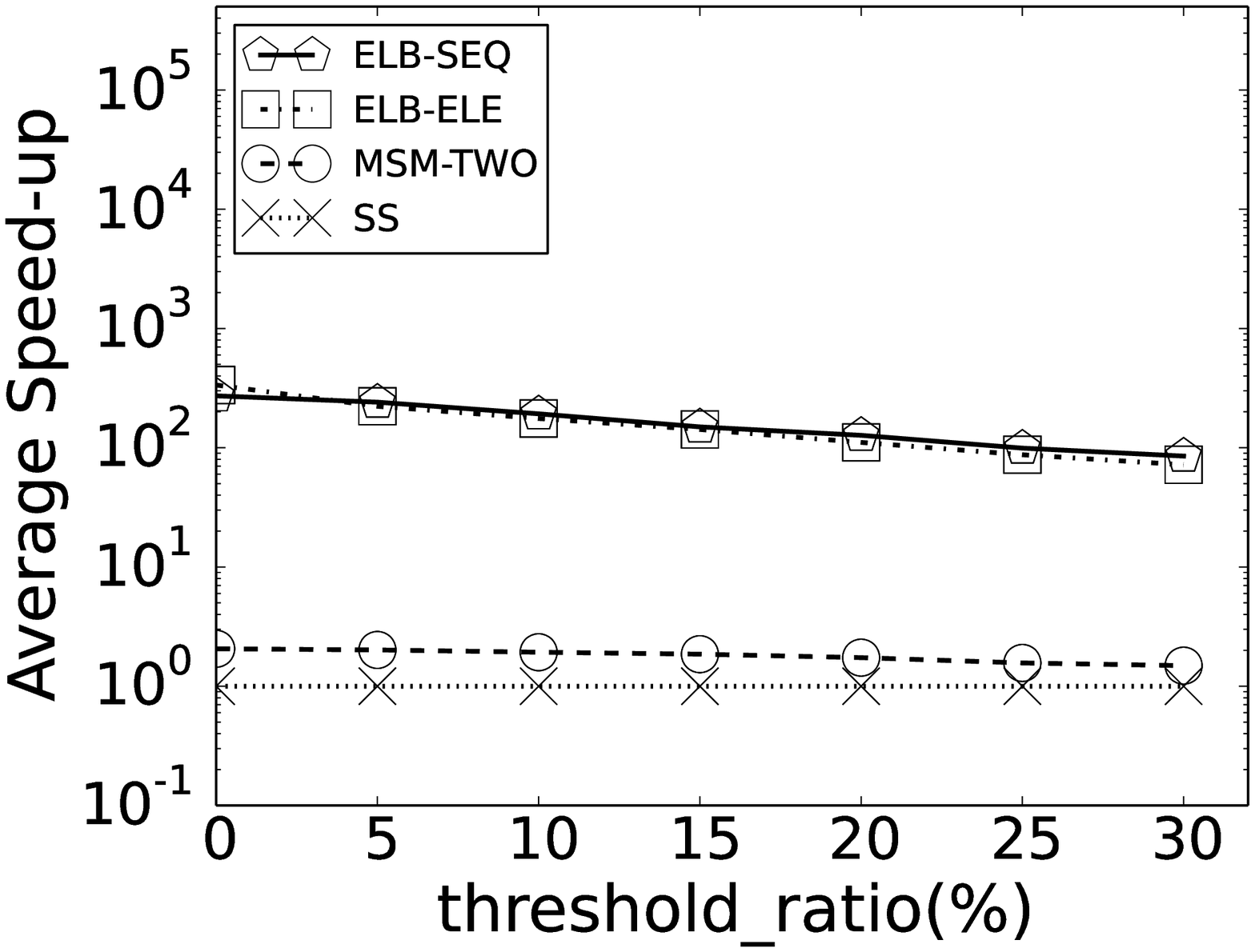}
		\caption[]%
		{ UCR\_ECG}
	\end{subfigure}
	\hfill
	\begin{subfigure}[b]{0.24\textwidth}   
		\centering 
		\includegraphics[width=1.05\textwidth]{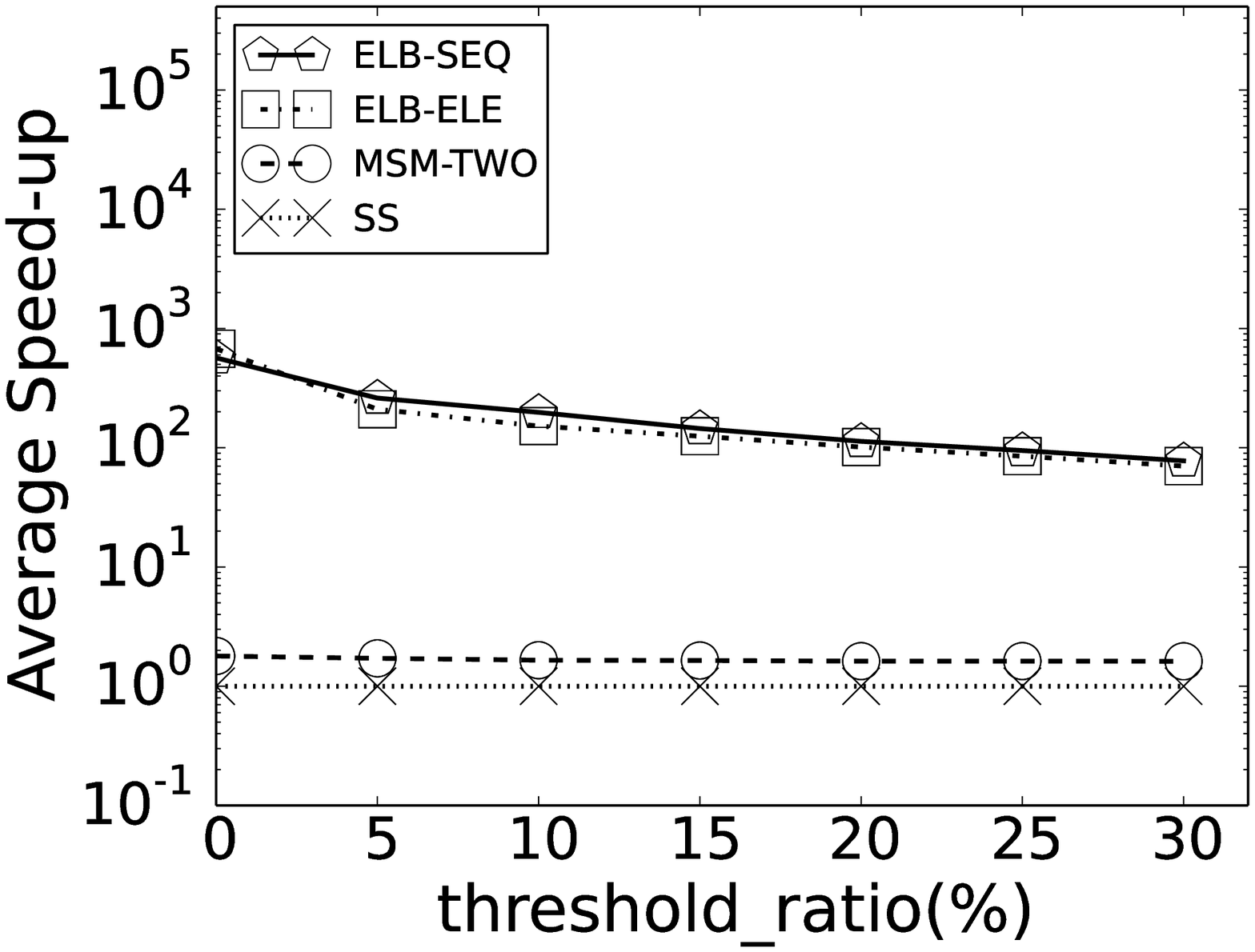}
		\caption[]%
		{UCR\_MALLAT}
	\end{subfigure}
	\caption{Speedup vs. \textit{threshold\_ratio}}
	\label{fig:4_threshold}
\end{figure}

The result on synthetic datasets is shown in Fig.~\ref{fig:4_threshold}. 
The performances of our two algorithms are very similar in synthetic datasets.
Both ELB-ELE and ELB-SEQ outperforms MSM and SS by orders of magnitude.
As the threshold gets larger, the speedups of ELB-ELE and ELB-SEQ decrease slightly.
Nevertheless, our algorithms keep their advantage over other approaches even though \textit{threshold\_ratio} increases to 30\%.

\subsubsection{Impact of Pattern Occurrence Probability} 
\label{sub:comparision_of_matching_ratio}
In this section, we further examine the performance by varying
the pattern occurrence probability.
When the probability becomes lower, more windows are filtered out in the pruning phase.
In contrast, when the probability becomes higher, more windows enter the post-processing phase.
A good approach should be robust to these situations.

\begin{figure}[tb]
	\centering
	\begin{subfigure}[b]{0.24\textwidth}
		\centering
		\includegraphics[width=1.05\textwidth]{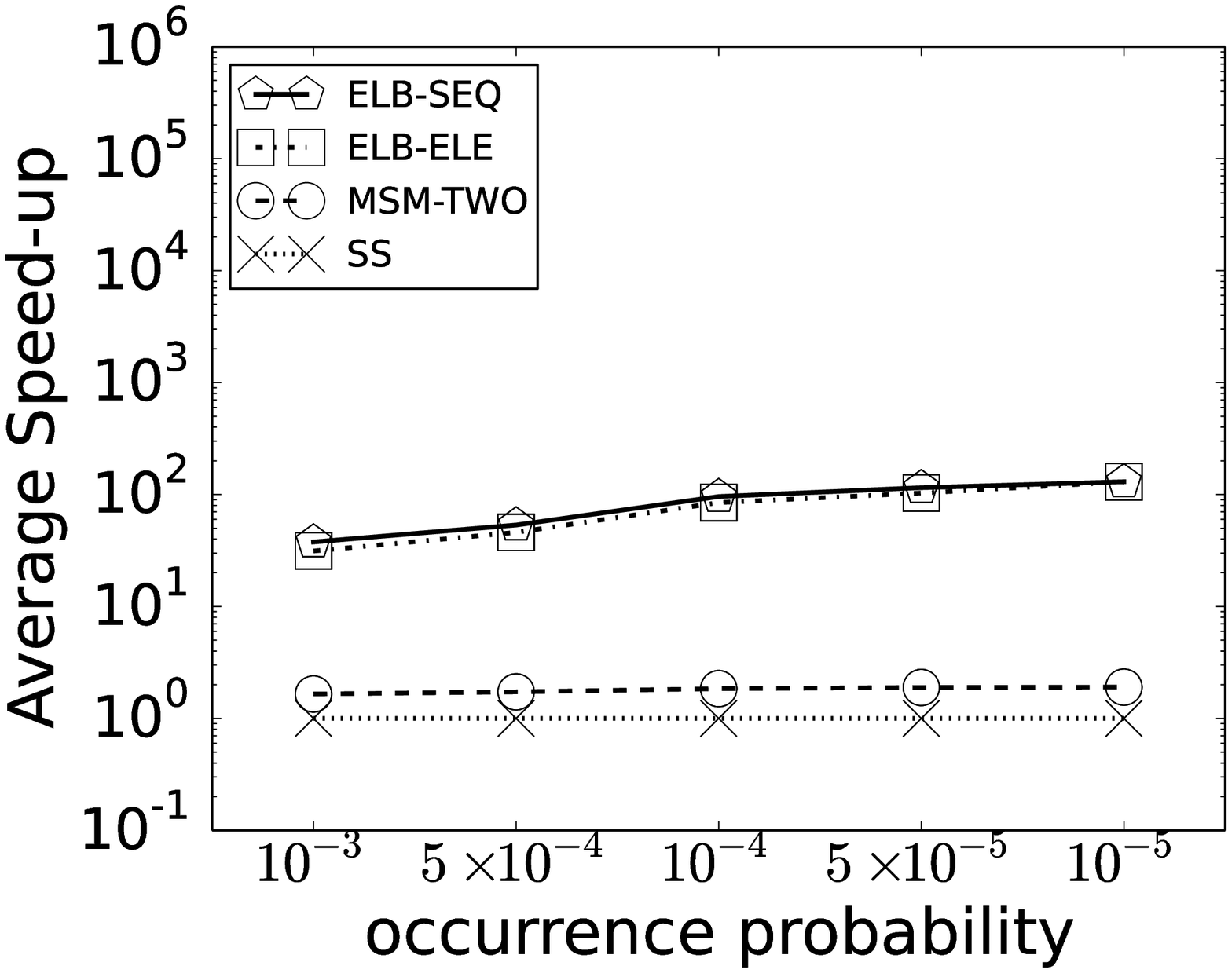}
		\caption[Network2]%
		{UCR\_Straw}
	\end{subfigure}
	\hfill
	\begin{subfigure}[b]{0.24\textwidth}  
		\centering 
		\includegraphics[width=1.05\textwidth]{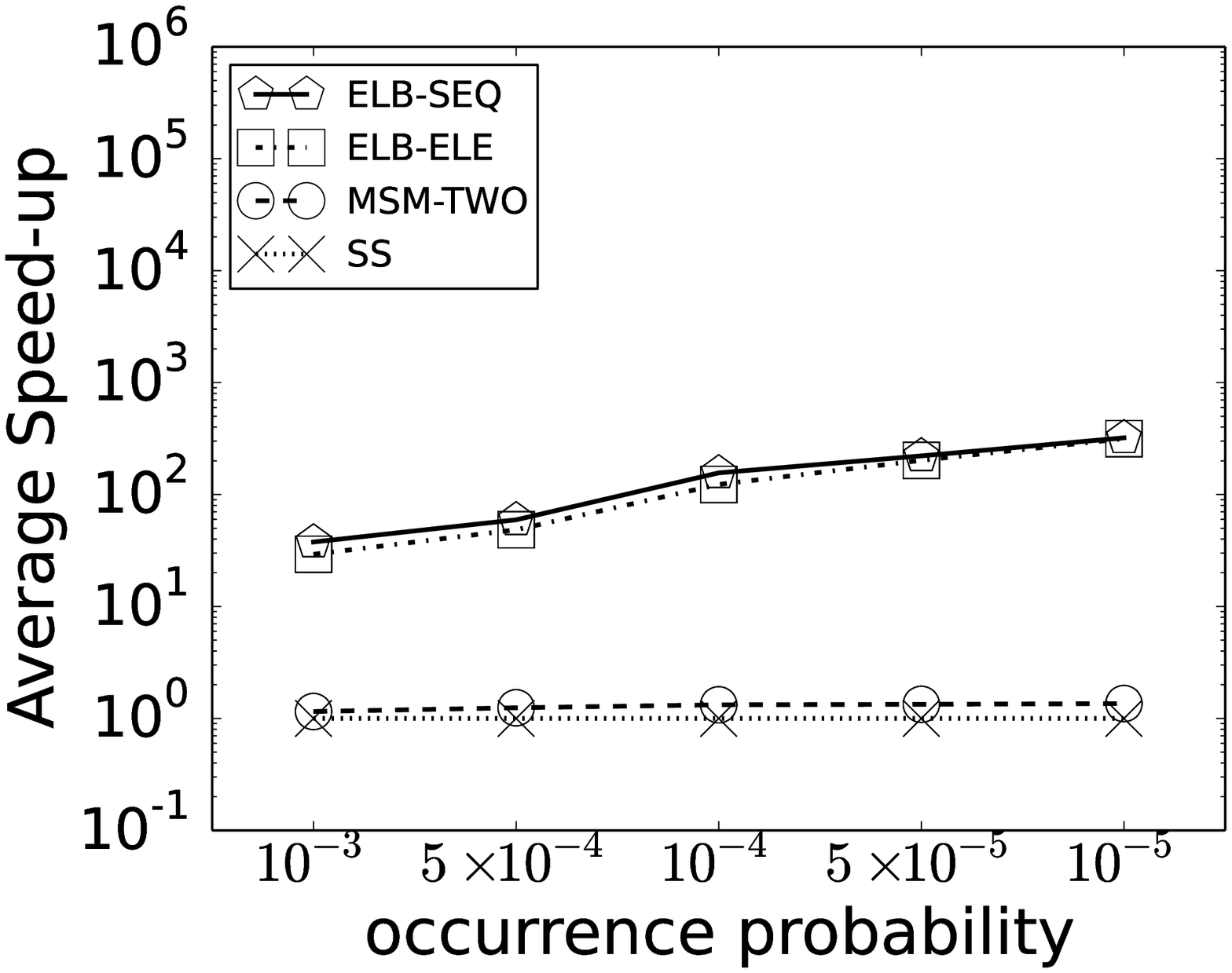}
		\caption[]%
		{UCR\_Meat}    
	\end{subfigure}
	\begin{subfigure}[b]{0.24\textwidth}   
		\centering 
		\includegraphics[width=1.05\textwidth]{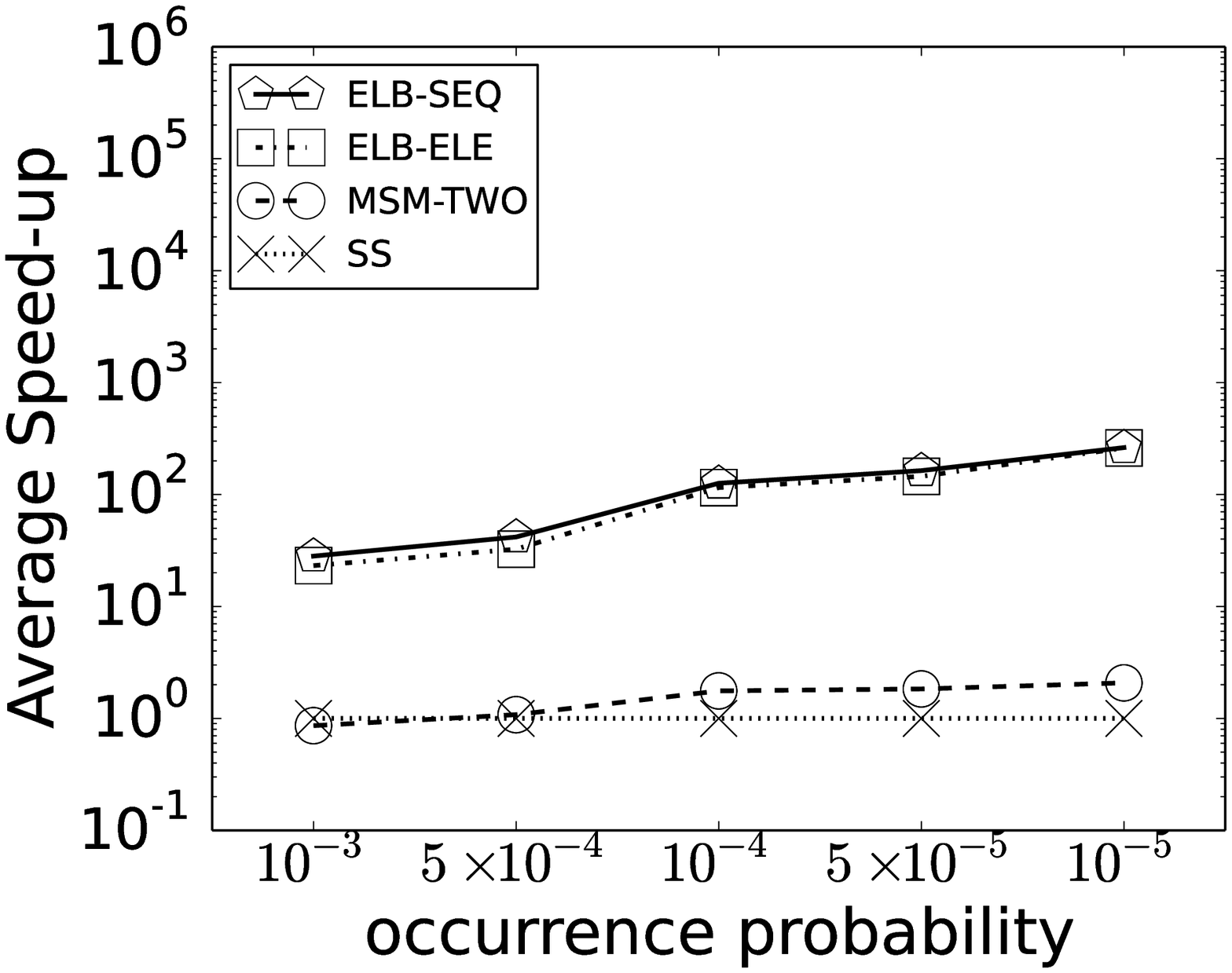}
		\caption[]%
		{UCR\_ECG}
	\end{subfigure}
	\hfill
	\begin{subfigure}[b]{0.24\textwidth}   
		\centering 
		\includegraphics[width=1.05\textwidth]{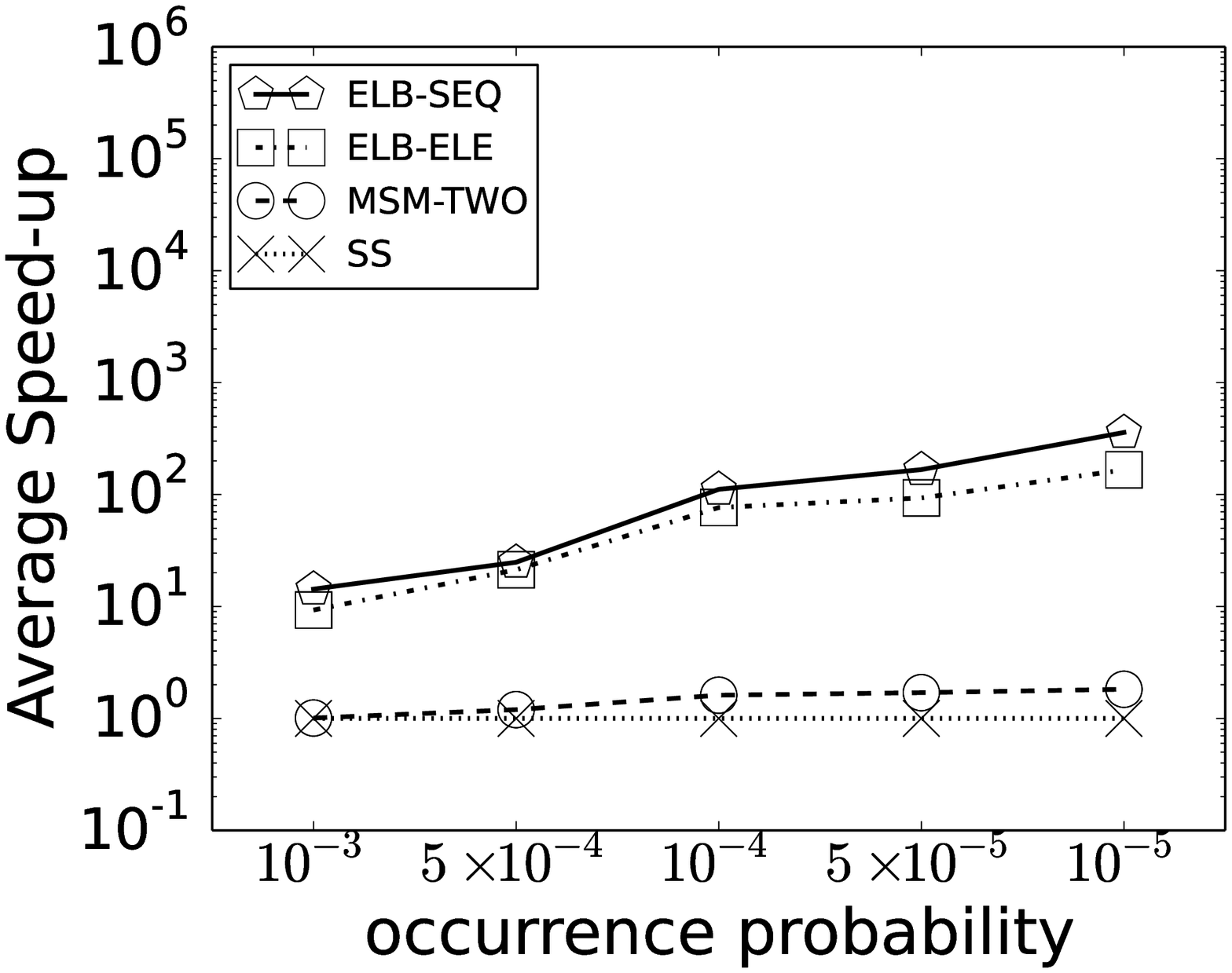}
		\caption[]%
		{UCR\_MALLAT}
	\end{subfigure}
	\caption{Speedup vs. pattern occurrence probability.}
	\label{fig:6_pattern_rate}
\end{figure}

We perform this experiment on synthetic datasets
and vary the occurrence probability over 
$ \{10^{-3}, 5 \times 10^{-4}, 10^{-4}, 5\times 10^{-5},10^{-5} \}$.
The largest probability is set to $ 10^{-3} $ since in this case, the stream of MALLAT, which has largest pattern length, has been filled up by embedded UCR time series.
As illustrated in Fig.~\ref{fig:6_pattern_rate}, 
Our algorithms outperform MSM and SS in all examined probabilities. 
Furthermore, our algorithms show a larger speedup when the pattern occurrence probability becomes lower.
This experiment demonstrates the robustness of our algorithms over different occurrence probabilities.

\subsubsection{Impact of Block Size} 
\label{sub:performance_block_size}
The block size is an important parameter affecting the pruning power of our approach.
In this experiment, we investigate the effect of block size by comparing ELB-ELE,  ELB-SEQ and MSM on both synthetic and real-world datasets.
We vary the ratio of the block size to the pattern length from $1\%$ to $40\%$.
A ratio being larger than $50\%$ indicates that the entire pattern contains only one block, which makes ELE-SEQ meaningless.

\begin{figure}[!tp]
	\centering
	\begin{subfigure}[b]{0.24\textwidth}
		\centering
		\includegraphics[width=1.05\textwidth]{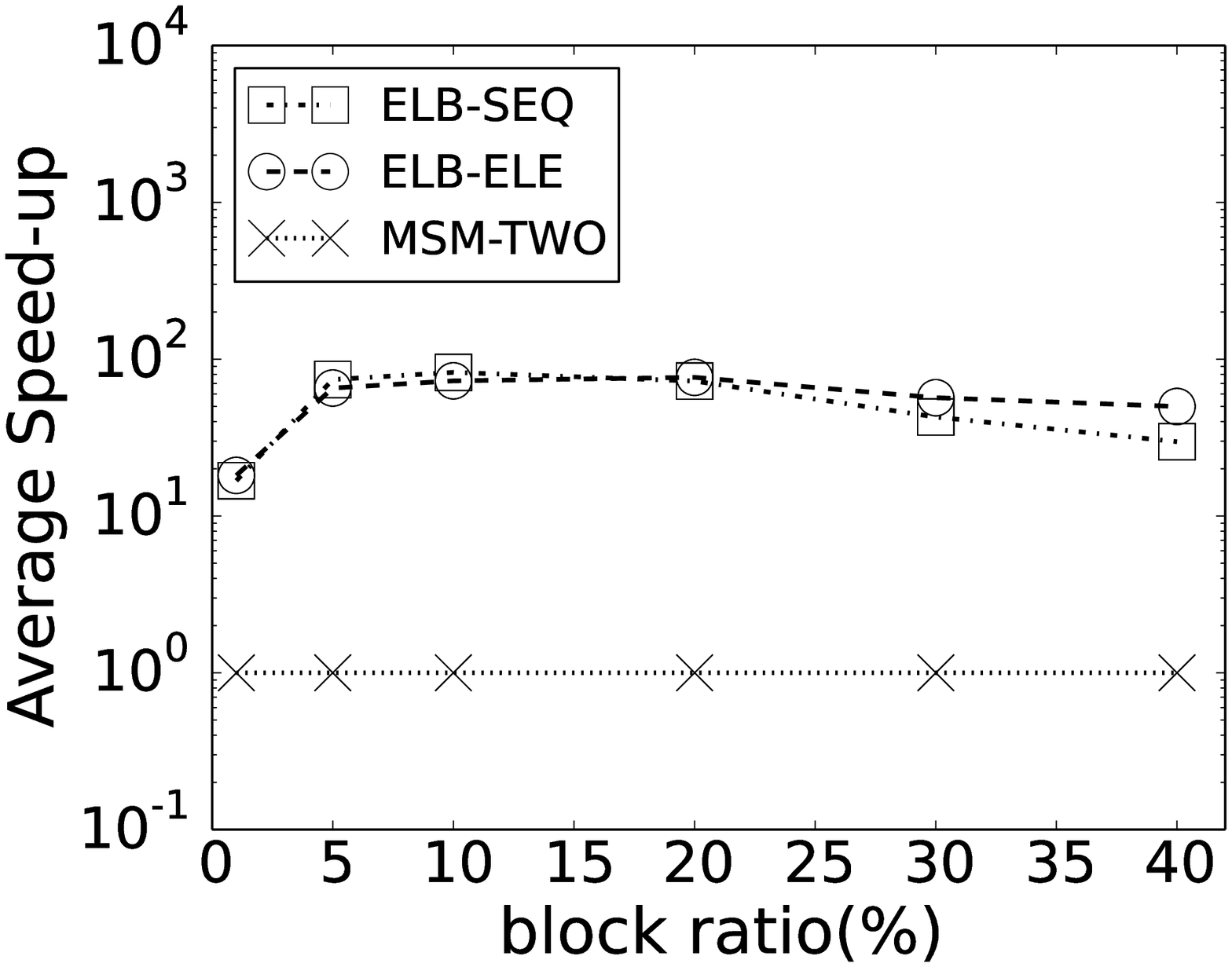}
		\caption[]%
		{ UCR\_Straw}
	\end{subfigure}
	\hfill
	\begin{subfigure}[b]{0.24\textwidth}
		\centering
		\includegraphics[width=1.05\textwidth]{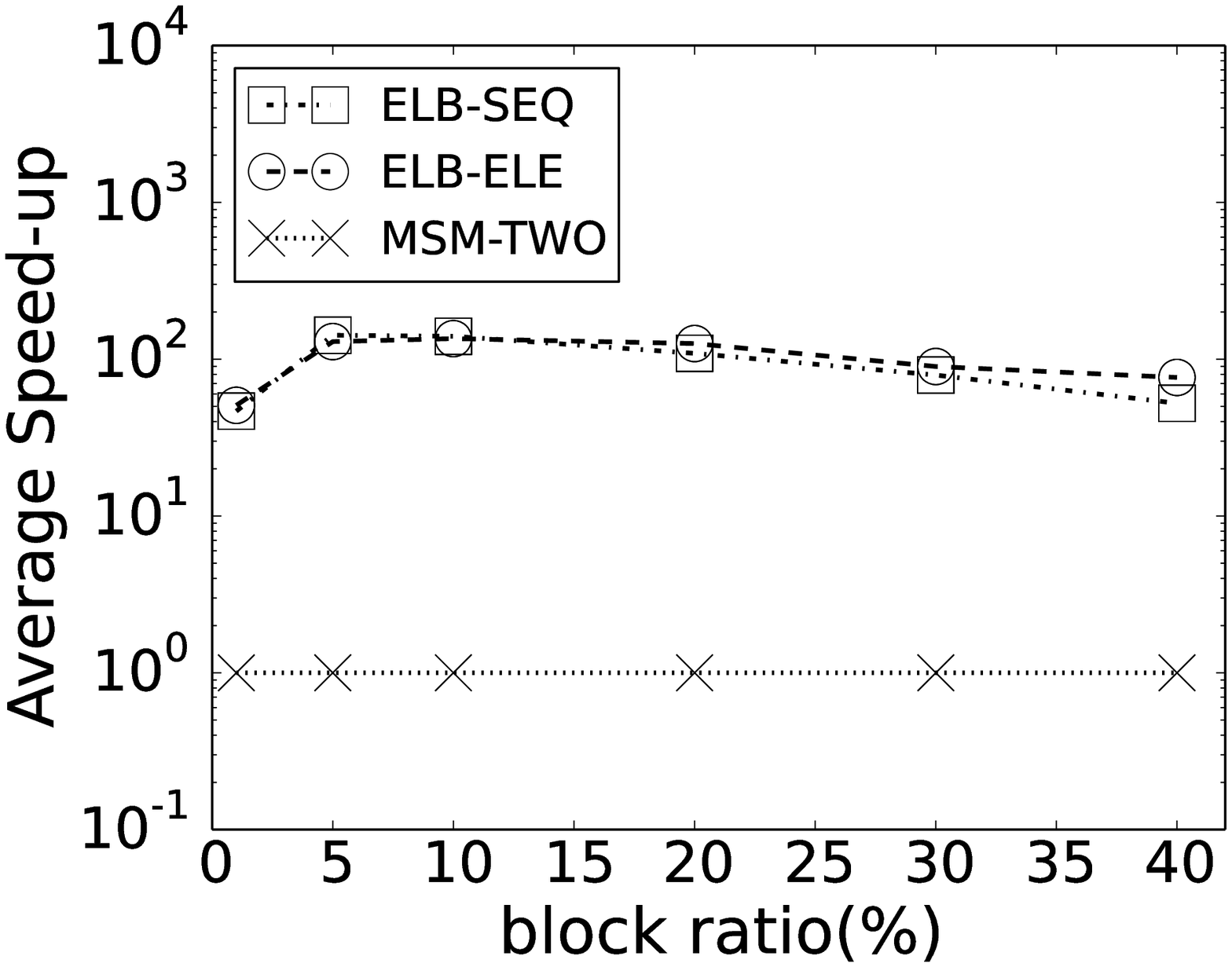}
		\caption[]%
		{ UCR\_Meat}
	\end{subfigure}
	\begin{subfigure}[b]{0.24\textwidth}   
		\centering 
		\includegraphics[width=1.05\textwidth]{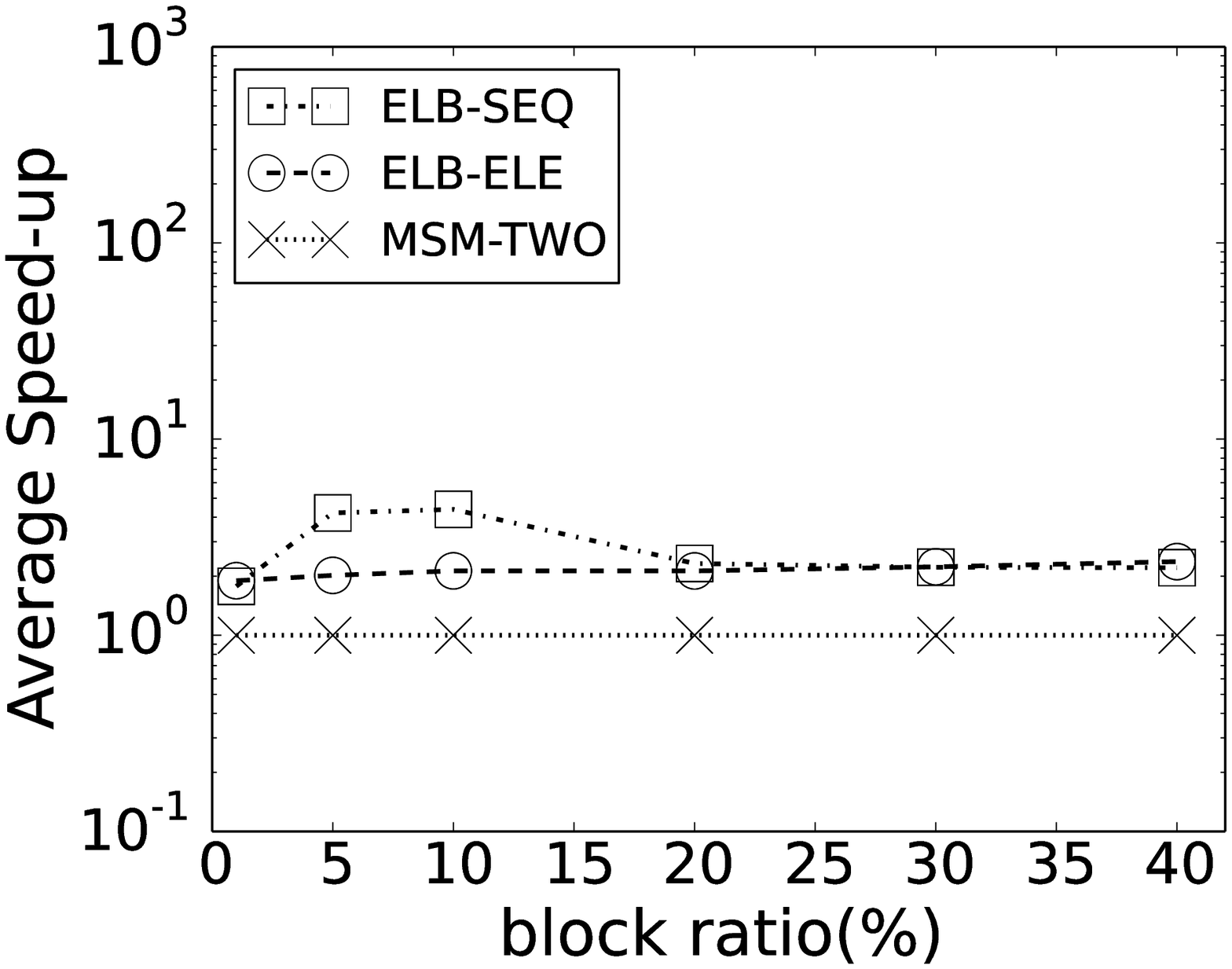}
		\caption[]%
		{ wind direction
	}
	\end{subfigure}
	\hfill
	\begin{subfigure}[b]{0.24\textwidth}   
		\centering 
		\includegraphics[width=1.05\textwidth]{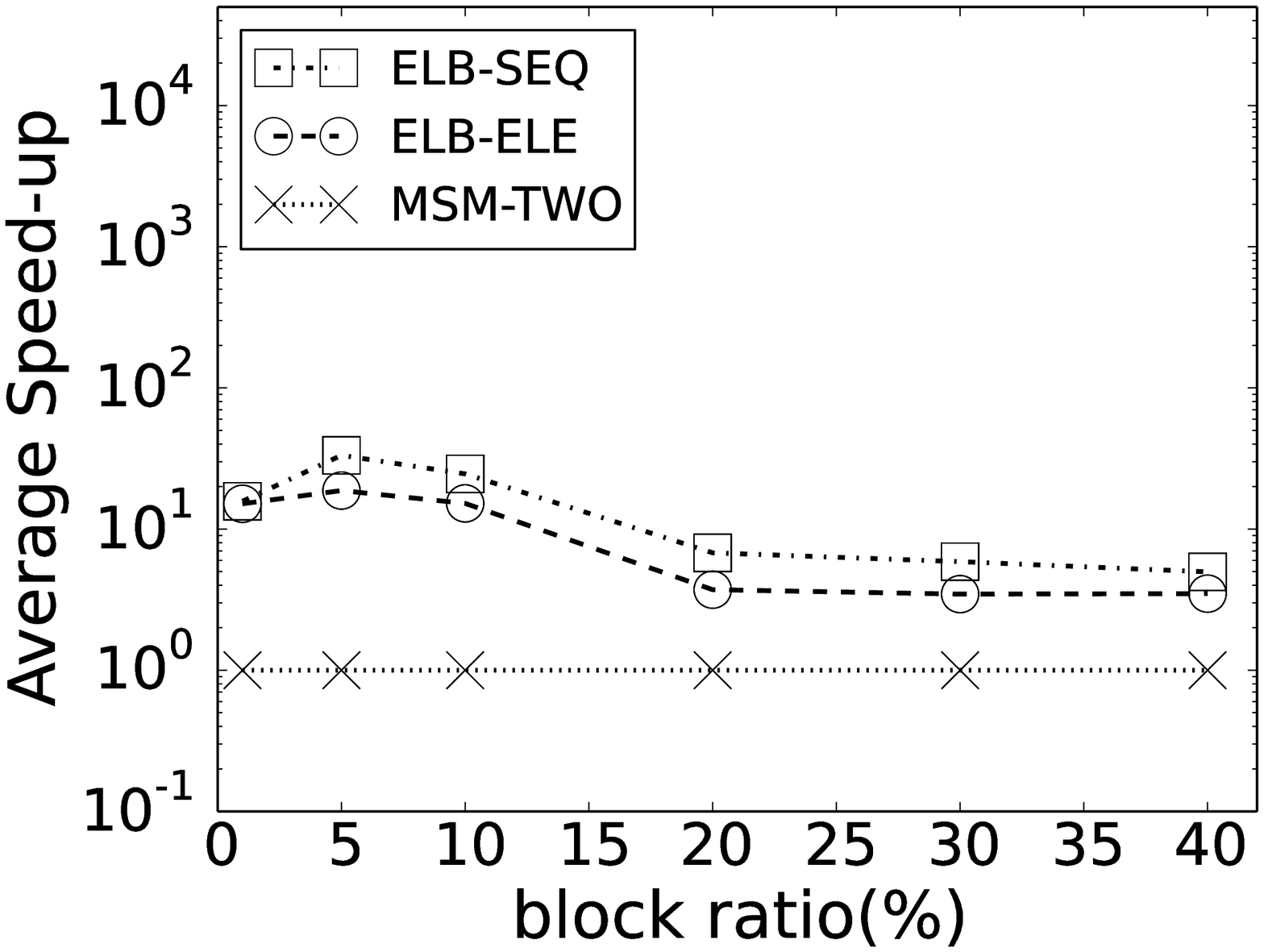}
		\caption[]%
		{ generator speed}
	\end{subfigure}
	\caption{Speedup vs. \textit{block\_ratio}.}
	\label{fig:1_window_block}
\end{figure}

Figure~\ref{fig:1_window_block} shows the experimental results on some representative synthetic and real-world datasets while the rest are consistent. 
A too small or too large block size results in performance degradation.
In detail, 
a smaller block size leads to a tighter bound for each block which improves the pruning effectiveness.
Nevertheless,  a small block size, corresponding to a small sliding step, results in more block computation and higher cost in the pruning phase.
A larger block size may bring less block computation, but a looser bound meanwhile. 
The loose bound incurs degradation of the pruning effectiveness.
In practice, our algorithms achieve the optimal performance when the block ratio is about $5\%$ to $10\%$.

%% file: sec6-conclusion.tex
\section{Conclusion}\label{sec:conclusion}
In this paper, we propose a new problem, called ``consecutive subpatterns matching'', 
which allows users to specify a pattern containing a list of consecutive subpatterns with different distance thresholds. 
We present a novel ELB representation to prune sliding windows efficiently under all $ L_p $-Norms. 
We conduct extensive experiments on both synthetic and real-world datasets to illustrate that our algorithm outperforms the baseline solution and prior-arts. 

%% file: 129.bbl
\begin{thebibliography}{10}

\bibitem{agrawal_efficient_1993}
R.~Agrawal, C.~Faloutsos, and A.~Swami.
\newblock Efficient similarity search in sequence databases.
\newblock In {\em Foundations of {Data} {Organization} and {Algorithms}}, pages
  69--84. Springer, Berlin, Heidelberg, Oct. 1993.
\newblock DOI: 10.1007/3-540-57301-1\_5.

\bibitem{begum_rare_2014}
N.~Begum and E.~Keogh.
\newblock Rare time series motif discovery from unbounded streams.
\newblock {\em PVLDB}, 8(2):149--160, 2014.

\bibitem{berndt_using_1994}
D.~J. Berndt and J.~Clifford.
\newblock Using {Dynamic} {Time} {Warping} to {Find} {Patterns} in {Time}
  {Series}.
\newblock In {\em {KDD} workshop}, volume~10, pages 359--370, 1994.

\bibitem{branlard_wind_2009}
E.~Branlard.
\newblock Wind energy: {On} the statistics of gusts and their propagation
  through a wind farm.
\newblock {\em ECN-Wind-Memo-09}, 5, 2009.

\bibitem{faloutsos_fast_1994}
C.~Faloutsos, M.~Ranganathan, and Y.~Manolopoulos.
\newblock Fast {Subsequence} {Matching} in {Time}-series {Databases}.
\newblock In {\em {SIGMOD}}, pages 419--429. ACM, 1994.

\bibitem{jensen_time_2017}
S.~K. Jensen, T.~B. Pedersen, and C.~Thomsen.
\newblock Time {Series} {Management} {Systems}: {A} {Survey}.
\newblock {\em TKDE}, PP(99):1--1, 2017.

\bibitem{keogh_welcome_nodate}
E.~Keogh.
\newblock Welcome to the {UCR} {Time} {Series} {Classification}/{Clustering}
  {Page}: www.cs.ucr.edu/{\textasciitilde}eamonn/time\_series\_data.

\bibitem{keogh_exact_2002}
E.~Keogh.
\newblock Exact {Indexing} of {Dynamic} {Time} {Warping}.
\newblock In {\em {PVLDB}}, pages 406--417, Hong Kong, China, 2002.

\bibitem{kotsifakos_subsequence_2011}
A.~Kotsifakos, P.~Papapetrou, J.~Hollmén, and D.~Gunopulos.
\newblock A subsequence matching with gaps-range-tolerances framework: a
  query-by-humming application.
\newblock {\em PVLDB}, 4(11):761--771, 2011.

\bibitem{lian_multiscale_2009}
X.~Lian, L.~Chen, J.~X. Yu, J.~Han, and J.~Ma.
\newblock Multiscale representations for fast pattern matching in stream time
  series.
\newblock {\em TKDE}, 21(4):568--581, 2009.

\bibitem{lian_similarity_2007}
X.~Lian, L.~Chen, J.~X. Yu, G.~Wang, and G.~Yu.
\newblock Similarity {Match} {Over} {High} {Speed} {Time}-{Series} {Streams}.
\newblock In {\em {ICDE}}, pages 1086--1095. IEEE, Apr. 2007.

\bibitem{lim_similar_2008}
H.-S. Lim, K.-Y. Whang, and Y.-S. Moon.
\newblock Similar sequence matching supporting variable-length and
  variable-tolerance continuous queries on time-series data stream.
\newblock {\em Information Sciences}, 178(6):1461--1478, 2008.

\bibitem{lim_using_2006}
S.-H. Lim, H.-J. Park, and S.-W. Kim.
\newblock Using {Multiple} {Indexes} for {Efficient} {Subsequence} {Matching}
  in {Time}-{Series} {Databases}.
\newblock In {\em {DASFAA}}, pages 65--79. Springer Berlin Heidelberg, Apr.
  2006.

\bibitem{loh_subsequence_2004}
W.-K. Loh, S.-W. Kim, and K.-Y. Whang.
\newblock A subsequence matching algorithm that supports normalization
  transform in time-series databases.
\newblock {\em DMKD}, 9(1):5--28, 2004.

\bibitem{luo_piecewise_2015}
G.~Luo, K.~Yi, S.~W. Cheng, Z.~Li, W.~Fan, C.~He, and Y.~Mu.
\newblock Piecewise linear approximation of streaming time series data with
  max-error guarantees.
\newblock In {\em 2015 {IEEE} 31st {International} {Conference} on {Data}
  {Engineering}}, pages 173--184, Apr. 2015.

\bibitem{moon_general_2002}
Y.-S. Moon, K.-Y. Whang, and W.-S. Han.
\newblock General match: a subsequence matching method in time-series databases
  based on generalized windows.
\newblock In {\em {SIGMOD}}, pages 382--393. ACM, 2002.

\bibitem{pace_lidar-based_2012}
A.~Pace, K.~Johnson, and A.~Wright.
\newblock Lidar-based extreme event control to prevent wind turbine overspeed.
\newblock In {\em 51st {AIAA} {Aerospace} {Sciences} {Meeting} including the
  {New} {Horizons} {Forum} and {Aerospace} {Exposition}}, page 315, 2012.

\bibitem{sun_matching_2009}
H.~Sun, K.~Deng, F.~Meng, and J.~Liu.
\newblock Matching {Stream} {Patterns} of {Various} {Lengths} and {Tolerances}.
\newblock In {\em {CIKM}}, pages 1477--1480. ACM, 2009.

\bibitem{vlachos_discovering_2002}
M.~Vlachos, G.~Kollios, and D.~Gunopulos.
\newblock Discovering similar multidimensional trajectories.
\newblock In {\em {ICDE}}, pages 673--684. IEEE, 2002.

\bibitem{wang_data-adaptive_2013}
Y.~Wang, P.~Wang, J.~Pei, W.~Wang, and S.~Huang.
\newblock A {Data}-adaptive and {Dynamic} {Segmentation} {Index} for {Whole}
  {Matching} on {Time} {Series}.
\newblock {\em PVLDB}, 6(10):793--804, Aug. 2013.

\bibitem{wei_atomic_2005}
L.~Wei, E.~Keogh, H.~Van~Herle, and A.~Mafra-Neto.
\newblock Atomic wedgie: efficient query filtering for streaming time series.
\newblock In {\em {ICDM}}, pages 8--pp. IEEE, 2005.

\bibitem{wu_online_2004}
H.~Wu, B.~Salzberg, and D.~Zhang.
\newblock Online event-driven subsequence matching over financial data streams.
\newblock In {\em {SIGMOD}}, pages 23--34. ACM, 2004.

\bibitem{yi_fast_2000}
B.-K. Yi and C.~Faloutsos.
\newblock Fast {Time} {Sequence} {Indexing} for {Arbitrary} {Lp} {Norms}.
\newblock In {\em {PVLDB}}, pages 385--394. Morgan Kaufmann Publishers Inc.,
  2000.

\bibitem{zhao_adaptive_2014}
J.~Zhao, K.~Liu, W.~Wang, and Y.~Liu.
\newblock Adaptive fuzzy clustering based anomaly data detection in energy
  system of steel industry.
\newblock {\em Information Sciences}, 259(Supplement C):335--345, Feb. 2014.

\bibitem{zhu_efficient_2003}
Y.~Zhu and D.~Shasha.
\newblock Efficient elastic burst detection in data streams.
\newblock In {\em {SIGKDD}}, pages 336--345. ACM, 2003.

\end{thebibliography}
